\documentclass[aps,pra,twocolumn]{revtex4-1}

\usepackage{graphicx}
\usepackage{dcolumn}
\usepackage{amssymb}
\usepackage{amsmath}
\usepackage{bbold}
\usepackage{bm}
\usepackage{color}
\usepackage{physics}
\usepackage{soul}

\usepackage[colorlinks,urlcolor=blue,citecolor=blue,linkcolor=blue]{hyperref}
\usepackage{comment}
\usepackage{cleveref}
\usepackage{mathtools}
\usepackage{mathrsfs}

\newcommand{\sch}{Schr{\"o}dinger }

\newcommand{\0}{\mathbf{0}}

\newcommand{\q}{\mathbf{q}}
\newcommand{\Q}{\mathbf{Q}}
\renewcommand{\k}{\mathbf{k}}

\makeatletter
\newcommand{\subalign}[1]{%
  \vcenter{%
    \Let@ \restore@math@cr \default@tag
    \baselineskip\fontdimen10 \scriptfont\tw@
    \advance\baselineskip\fontdimen12 \scriptfont\tw@
    \lineskip\thr@@\fontdimen8 \scriptfont\thr@@
    \lineskiplimit\lineskip
    \ialign{\hfil$\m@th\scriptstyle##$&$\m@th\scriptstyle{}##$\hfil\crcr
      #1\crcr
    }%
  }%
}
\makeatother

\begin{document}

\title{Coupled-channel approach to scattering of hybrid excitons}

\author{Yasufumi Nakano}
\affiliation{School of Physics and Astronomy, Monash University, Victoria 3800, Australia}

\author{Meera M. Parish}
\affiliation{School of Physics and Astronomy, Monash University, Victoria 3800, Australia}

\author{Jesper Levinsen}
\affiliation{School of Physics and Astronomy, Monash University, Victoria 3800, Australia}

\date{\today}

\begin{abstract}
We consider the interactions of hybrid excitons in a two-dimensional semiconductor bilayer, where spatially direct and indirect excitons are hybridized by interlayer charge-carrier tunneling. Starting from a microscopic electron-hole description, we construct realistic pseudopotentials for exciton-exciton interactions and use them as inputs to a coupled-channel scattering integral equation. This enables non-perturbative calculations of hybrid-exciton scattering beyond standard perturbative theories, and highlights the importance of energy-dependent scattering and channel mixing. In particular, we show that the interaction strength of hybrid excitons exhibits a rapid growth with increasing energy, which we find is inherited from their indirect-exciton component. We further demonstrate that dielectric screening affects the direct and indirect channels in distinct ways, leading to markedly different interaction strengths across experimentally relevant dielectric environments. Finally, we show that the hybrid-exciton scattering strength can be electrically tuned via the Stark shift, which controls the direct-indirect detuning and hence the hybridization of the two exciton modes.
\end{abstract}

\maketitle

\section{Introduction}
\begin{figure}[tp]
    \begin{minipage}{0.48\textwidth}
    \centering
    \includegraphics[width=\textwidth]{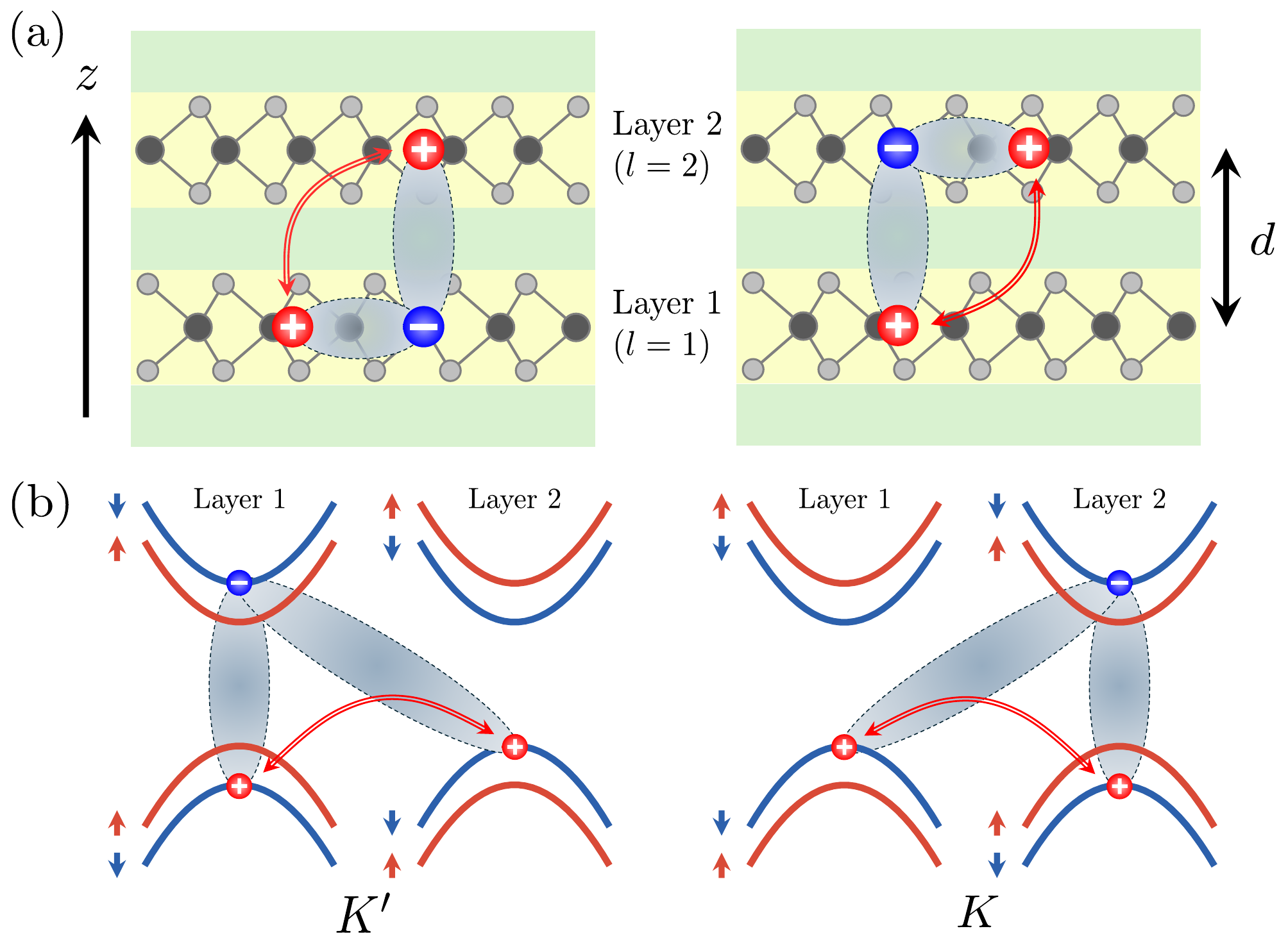}
    \end{minipage}
    \caption{(a) Schematic of a hybrid exciton in a naturally stacked 2H MoS$_2$ homobilayer. An electron (blue) and a hole (red) form either a DX or an IX, hybridized via interlayer hole tunneling. The left and right panels show the configurations in the $K'$ and $K$ valleys, respectively, which correspond to opposite orientations of the IX out-of-plane dipole moment. (b) Corresponding band structure in the $K'$ (left) and $K$ (right) valleys.}
    \label{fig:schematic}
\end{figure}

In two-dimensional (2D) semiconductors, Coulomb-bound electron-hole pairs form excitons with strong optical response and appreciable interactions~\cite{Wang2018}. The prospect of controlling exciton interactions in situ is of great interest, since their strength and range set key scales for exciton dynamics~\cite{Moody2016,Perea2022,Li2025} and are responsible for the emergence of interaction-driven phases~\cite{Keldysh2024,Butov2002,Eisenstein2004,Xiong2023}. A promising route to achieving highly tunable exciton-exciton interactions is offered by bilayers of atomically thin transition metal dichalcogenides (TMDs) hosting hybrid excitons~\cite{Deilmann2018,Gerber2019,Leisgang2020,Lorchat2021,Lopriore2025,Federolf2025}. Here, as illustrated in Fig.~\ref{fig:schematic}, spatially direct (DX) and indirect (IX) excitons hybridize via charge-carrier tunneling between the two layers, creating hybrid exciton modes whose composition is adjustable. The key control parameter is the DX-IX detuning, i.e., the energy offset between the DX and IX resonances, which can be tuned by an out-of-plane applied electric field via the Stark shift. Varying this detuning redistributes the direct and indirect excitonic properties to the hybrid exciton modes, where the direct component can strongly couple to light, thus providing optical control, while the indirect component yields long-range dipole–dipole interactions. Additionally, there is a possibility of enhancing dipole–dipole interactions via the surrounding dielectric environment~\cite{Maslova2024}.

Exciton-exciton interactions in 2D semiconductors are often treated 
perturbatively using the lowest-order Born approximation~\cite{Ciuti1998,Tassone1999,Glazov2009}. While this approach accounts for the electron and hole degrees of freedom within the exciton, it fails to capture the non-perturbative behavior of exciton scattering at low collision energies relative to the exciton binding energy---the relevant regime for TMD monolayers~\cite{Bleu2020,Li2021c}. The situation is even more complex for hybrid excitons in bilayers, where the relevant excitonic eigenmodes are superpositions of direct and indirect excitons generated by interlayer charge-carrier tunneling. The scattering problem then becomes intrinsically multichannel, as collisions between hybrid excitons are mediated by virtual transitions into different underlying two-exciton configurations with distinct interaction potentials. Capturing such processes requires a non-perturbative theory that goes beyond a single-channel description of the interactions.

In this work, we develop a coupled-channel $T$-matrix approach for hybrid excitons, in which the exact two-body scattering between the physical hybrid eigenmodes is determined through a multichannel scattering formalism. Our theory naturally incorporates energy-dependent scattering between hybrid excitons, and unlike standard perturbative theories~\cite{Nalitov2014a,Maslova2024}, it is guaranteed to conserve probability in a given scattering process and to satisfy the universal low-energy scattering behavior in 2D systems~\cite{Landau2013quantum}. To connect this effective scattering theory to microscopic physics, we first construct bare DX and IX states from an electron-hole Hamiltonian and evaluate their interaction strengths in the Born approximation. We then use these microscopic results as inputs for effective pseudopotentials for excitons. This enables us to quantify how tunneling, dielectric screening, and the electrically tunable Stark shift act together to determine the scattering properties of hybrid excitons.

This paper is organized as follows. In Sec.~\ref{sec:microscopic description of excitons}, we present a microscopic description of direct and indirect excitons in the bilayer, and in Sec.~\ref{sec:microscopic description of exciton-exciton scattering} we discuss a microscopic description of their interactions. In Sec.~\ref{sec:hybrid exciton}, we then introduce an effective excitonic description of hybrid excitons, and we use this in Sec.~\ref{sec:multichannel approach to scattering of hybrid excitons} to develop a multichannel approach to hybrid-exciton scattering. We conclude in Sec.~\ref{sec:conclusions}.

\section{Microscopic description of excitons}
\label{sec:microscopic description of excitons}
\subsection{Microscopic model}
In this section, we construct the direct and indirect excitons in a two-dimensional bilayer within a microscopic approach based on the electron and hole degrees of freedom. To be concrete, we focus on the experimentally relevant case of a naturally stacked 2H (AA$'$) MoS$_2$ homobilayer~\cite{Leisgang2020,Lorchat2021,Lopriore2025,Federolf2025} illustrated in Fig.~\ref{fig:schematic}(a), and we choose material parameters that are typical of these experiments. Throughout this work, we assume spin-polarized electrons and holes (of relevance to excitons created from co-circularly polarized photons). According to the band structure shown in Fig.~\ref{fig:schematic}(b), we can therefore suppress the spin index of the charges, and describe the system in terms of conduction band electrons and valence band holes labelled by their layer index $l \in \{1,2\}$ and valley index $\xi \in \{K,K'\}$.  As also shown in Fig.~\ref{fig:schematic}(b), we focus on the $B$ direct excitons since these are typically closer in energy to the indirect excitons than $A$ excitons, and therefore have stronger hybridization.

Note that, in order to obtain distinct direct and indirect excitons at the microscopic level, in this section we do not explicitly include interlayer hole tunneling or an applied out-of-plane electric field. These effects, which lead to hybridization of the exciton modes, are instead included in the excitonic model in Sec.~\ref{sec:hybrid exciton}. There, hole tunneling is treated perturbatively in the unhybridized DX-IX basis, while the electric field enters through the Stark shift of the IX. The validity of this approach for the optically relevant hybrid exciton branches is checked by comparison with the microscopic coupled \sch equation.

The microscopic Hamiltonian takes the form
\begin{widetext}
\begin{align}
    &\hat{H}_\text{eh} = \sum_{\k}\sum_{l,\xi}\left[(\epsilon_{\k}^\text{e} \hat{e}^\dag_{\k,l\xi}\hat{e}_{\k,l\xi}+(\epsilon_{\k}^\text{h}+\delta^\text{h}_{l\xi})\hat{h}^\dag_{\k,l\xi}\hat{h}_{\k,l\xi})\right] \notag \\ 
    &+ \frac{1}{2}\sum_{\substack{\k\k'\q}}\sum_{l,l'}\sum_{\xi,\xi'} U_\q^{ll'}\left[ \hat{e}^\dag_{\k+\q,l\xi}\hat{e}^\dag_{\k'-\q,l'\xi'}\hat{e}_{\k',l'\xi'}\hat{e}_{\k,l\xi}+\hat{h}^\dag_{\k+\q,l\xi}\hat{h}^\dag_{\k'-\q,l'\xi'}\hat{h}_{\k',l'\xi'}\hat{h}_{\k,l\xi} -2\hat{e}^\dag_{\k+\q,l\xi}\hat{h}^\dag_{\k'-\q,l'\xi'}\hat{h}_{\k',l'\xi'}\hat{e}_{\k,l\xi} \right].
    \label{eq:electron-hole Hamiltonian}
\end{align}
\end{widetext}
Here, $\hat{e}_{\k,l\xi}$ and $\hat{h}_{\k,l\xi}$ denote the annihilation operators of the electron and the hole, respectively, with in-plane momentum $\k$, layer index $l$, and valley index $\xi$. The corresponding electron and hole masses are denoted by $m_\text{e}$ and $m_\text{h}$. We use units in which $\hbar$ and the area are set to unity. Measured from the bottom of the corresponding conduction band, the electron dispersion takes the form $\epsilon^\text{e}_{\k} = k^{2}/2m_\text{e}$. The kinetic contribution to the hole dispersion is similarly written as $\epsilon^\text{h}_{\k} = k^{2}/2m_\text{h}$. However, owing to the $180^\circ$ rotation between the two layers in the 2H-stacked bilayer, the ordering of the relevant valence bands depends on the layer and valley, as illustrated in Fig.~\ref{fig:schematic}(b). We account for this through the band-edge offset
\begin{equation}
    \delta^\text{h}_{l\xi} =
    \begin{cases}
    \delta_\text{h}, &(l,\xi)=(1,K)\, \text{or}\, (2,K'),\\
    0, &(l,\xi)=(1,K')\, \text{or}\, (2,K),
    \end{cases}
    \label{eq:band-edge offset}
\end{equation}
where $\delta_{\text{h}}<0$ denotes the energy separation between the two relevant hole branches.

The charges interact via the standard Rytova-Keldysh potential~\cite{Rytova1967,Keldysh1979}, appropriately generalized to a bilayer~\cite{Asriyan2019,Semina2019}
\begin{subequations}
    \label{eq:bilayer Keldysh potential}
\begin{align}
    U^{11(22)}_{\q} &\equiv U_{\q} = \frac{2\pi e^{2}}{\kappa q}\frac{1+q\rho_{0}(1-e^{-2qd})}{(1+q\rho_{0})^{2}-q^{2}\rho^{2}_{0}e^{-2qd}}, \label{eq:intralayer Keldysh potential} \\
    U^{12(21)}_{\q} &\equiv U^{d}_{\q} = \frac{2\pi e^{2}}{\kappa q}\frac{e^{-qd}}{(1+q\rho_{0})^{2}-q^{2}\rho^{2}_{0}e^{-2qd}}.
     \label{eq:interlayer Keldysh potential}
\end{align}
\end{subequations}
Using the correspondence between the layer index $l$ in the electron/hole operators and in the potential of Eq.~\eqref{eq:electron-hole Hamiltonian}, we define $U^{11}_{\q} = U^{22}_{\q} \equiv U_{\q}$ as the intralayer Keldysh potential and $U^{12}_{\q} = U^{21}_{\q} \equiv U^{d}_{\q}$ as the interlayer Keldysh potential. In Eq.~\eqref{eq:bilayer Keldysh potential}, $e$ denotes the elementary charge, $\kappa$ denotes the environmental dielectric constant, and $d$ denotes the interlayer separation. We take $\kappa=1$ for vacuum and $\kappa=3.76$ for hBN encapsulation~\cite{Laturia2018}. The dielectric screening length is set by $\rho_{0} = 2\pi\chi_\text{2D}/\kappa$, with $\chi_\text{2D}$ denoting the in-plane polarizability of the layers. For convenience, we further introduce the effective 2D Bohr radius $a_{0} = \kappa/2\mu e^{2}$, where $\mu=m_\text{e}m_\text{h}/(m_\text{e}+m_\text{h})$ is the electron-hole reduced mass.  In the following, we suppress any explicit $\kappa$-dependence of $\rho_{0}$ and $a_{0}$ unless otherwise noted. The physical values of the parameters are summarized in Table~\ref{tab:parameters}.

\begin{table}[t]
    \centering
    \renewcommand{\arraystretch}{1.2}
    \setlength{\tabcolsep}{6pt}
    \begin{tabular}{lcc}
        \hline\hline
        Parameter & $\kappa=1$ & $\kappa=3.76$ \\
        \hline
        $a_{0}$ (nm) & 0.106 & 0.398 \\
        $\rho_0$ (nm) & 4.23 & 1.12 \\
        $a_\text{X}$ (nm) & 0.576 & 0.889 \\
        $\varepsilon_\text{X}$ (meV) & 460 & 192 \\
        \hline\hline
    \end{tabular}
    \caption{Relevant length and energy scales used in the present work for vacuum $\kappa=1$ and hBN encapsulation $\kappa=3.76$. Here, $a_{0}$ denotes the effective 2D Bohr radius, $\rho_{0}$ the dielectric screening length, $a_\text{X}$ the effective DX exciton size, and $\varepsilon_\text{X}$ the DX binding energy. In all cases, we take equal masses for the electron and hole as is approximately the case in homobilayer MoS$_2$ (see, e.g., Refs.~\cite{Berkelbach2013,Kylanpaa2015,Gerber2019}): $m_\text{e} = m_\text{h} = 0.5m_{0}$, where $m_{0}$ is the bare electron mass. In solving the \sch equation in Eq.~\eqref{eq:schrodinger equation DX} to determine $a_\text{X}$ and $\varepsilon_\text{X}$, we also fix the layer separation to $d=6a^{(\kappa=1)}_{0}=0.64$\,nm and the polarizability to $\chi_\text{2D}=40a^{(\kappa=1)}_{0}/2\pi=0.67$\,nm.}
    \label{tab:parameters}
\end{table}

The screening length $\rho_{0}$ sets the momentum scale at which the electronic interactions cross over from being dominated by the dielectric environment to being dominated by the 2D polarizability of the layer. This is readily seen by considering the long-wavelength regime $q\rho_{0} \ll 1$ in Eq.~\eqref{eq:intralayer Keldysh potential}, where the interaction reduces to the usual 2D Coulomb potential
\begin{equation}
    U_{\q} \simeq \frac{2\pi e^{2}}{\kappa q}.
\end{equation}
On the other hand, in the short-wavelength regime where $q\rho_{0} \gg 1$, one obtains
\begin{equation}
    U_{\q} \simeq \frac{2\pi e^{2}}{\kappa q^{2} \rho_{0}} = \frac{e^{2}}{\chi_\text{2D} q^{2}}.
\end{equation}
Thus, increasing $\kappa$ primarily suppresses the long-range (small-$q$) Coulomb tail, while the short-range (large-$q$) interaction is largely set by $\chi_\text{2D}$. We note that the standard monolayer interaction potential is obtained by taking the limit $d \to \infty$ in Eq.~\eqref{eq:intralayer Keldysh potential}.

\begin{figure}[tp]
    \begin{minipage}{0.47\textwidth}
    \centering
    \includegraphics[width=\textwidth]{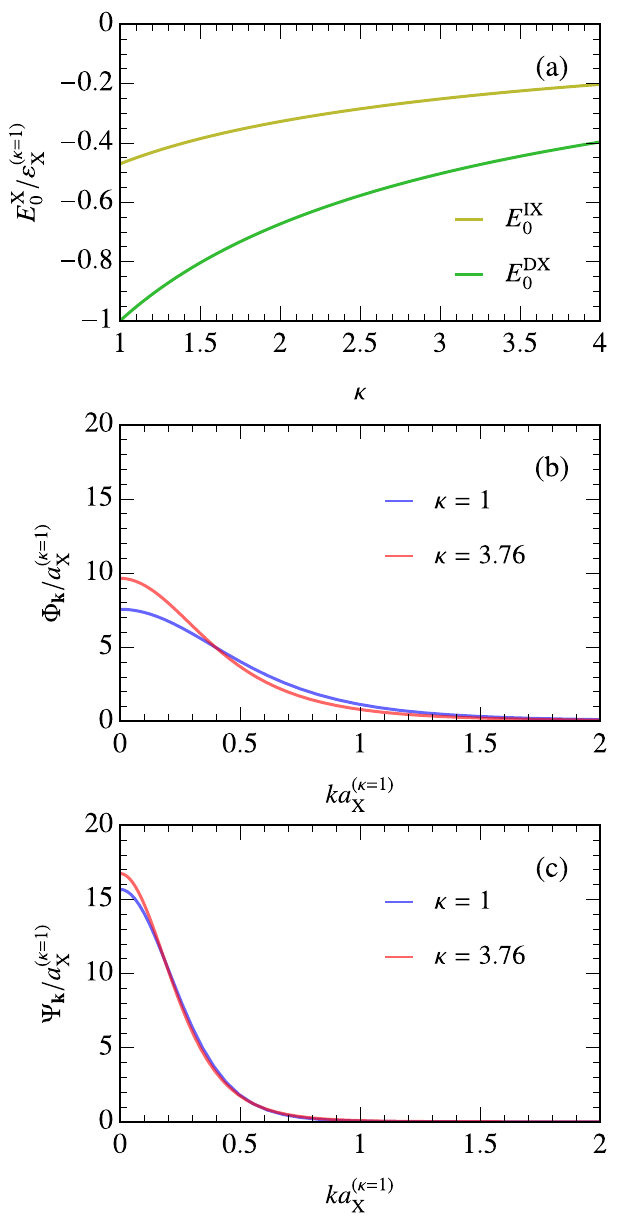}
    \end{minipage}
    \caption{(a) Ground-state energies of the DX (green) and IX (yellow) as a function of dielectric constant. (b,c) Ground-state wave functions for (b) the DX and (c) the IX. In (b,c), blue and red lines correspond to vacuum ($\kappa=1$) and hBN ($\kappa=3.76$), respectively. The values of all parameters, including the fixed reference scales $a_\text{X}^{(\kappa=1)}$ and $\varepsilon_\text{X}^{(\kappa=1)}$, are listed in Table~\ref{tab:parameters}.}
    \label{fig:Scheq}
\end{figure}

\subsection{Direct and indirect excitons}
To determine the energies and wave functions of a DX and an IX, we now consider the most general electron-hole wave functions at zero center-of-mass momentum
\begin{subequations}
    \label{eq:exciton state}
\begin{align}
    \ket{\Phi_{\text{DX},l\xi}} & =\sum_{\k} \phi_{\k} \hat{e}^{\dag}_{\k,l\xi}\hat{h}^{\dag}_{-\k,l\xi}\ket{0}, \label{eq:direct exciton state} \\
    \ket{\Psi_{\text{IX},l\xi}} & =\sum_{\k} \psi_{\k} \hat{e}^{\dag}_{\k,l\xi}\hat{h}^{\dag}_{-\k,\bar{l}\xi}\ket{0}, \label{eq:indirect exciton state}
\end{align}
\end{subequations}
where $\ket{0}$ denotes the vacuum state. The DX state involves an electron and hole residing in the same layer, while the IX state has the electron and hole in different layers. Accordingly, we use $\bar{l}$ to denote the layer index opposite to $l$. Since we consider the hybridization of a DX with an IX induced by hole tunneling (see Fig.~\ref{fig:schematic}), we only consider excitons formed by the electron and hole in the same valley, and therefore the electron and hole in the exciton states in Eq.~\eqref{eq:exciton state} have the same valley index. For the choice of excitons and spins in Fig.~\ref{fig:schematic}(b), the DX has $(l,\xi)=(1,K')$ or $(2,K)$ while the IX has $(l,\xi)=(1,K)$ or $(2,K')$.

Projecting the electron-hole Hamiltonian in Eq.~\eqref{eq:electron-hole Hamiltonian} onto the states in Eq.~\eqref{eq:exciton state} gives
\begin{subequations}
    \label{eq:full schrodinger equations}
\begin{align}
    \mathcal{E}_{\text{DX},l\xi}\phi_{\k} &= \left( \bar{\epsilon}_{\k} + \delta^\text{h}_{l\xi} \right)\phi_{\k} - \sum_{\k'} U_{\k-\k'}\phi_{\k'}, \label{eq:full schrodinger equation DX} \\ 
    \mathcal{E}_{\text{IX},l\xi}\psi_{\k} &= \left( \bar{\epsilon}_{\k} + \delta^\text{h}_{\bar{l}\xi} \right)\psi_{\k} - \sum_{\k'} U^{d}_{\k-\k'}\psi_{\k'}, 
    \label{eq:full schrodinger equation IX}
\end{align}
\end{subequations}
where $\bar\epsilon_\k=\epsilon_\k^\text{e}+\epsilon_\k^\text{h} = k^2/2\mu$ is the kinetic energy of the relative motion of the electron and hole. The terms $\delta^\text{h}_{l\xi}$ and $\delta^\text{h}_{\bar{l}\xi}$ are the electron-hole continuum thresholds of the DX and IX channels, respectively. The corresponding energies measured relative to these thresholds are
\begin{subequations}
    \label{eq:relative exciton energies}
\begin{align}
    E_\text{DX} &\equiv \mathcal{E}_{\text{DX},l\xi} - \delta^\text{h}_{l\xi}, \\ 
    E_\text{IX} &\equiv \mathcal{E}_{\text{IX},l\xi} - \delta^\text{h}_{\bar{l}\xi}.
\end{align}
\end{subequations}
We therefore obtain the momentum-space \sch equations for the relative motion
\begin{subequations}
    \label{eq:relative schrodinger equations}
\begin{align}
    E_\text{DX}\phi_{\k} &= \bar\epsilon_{\k}\phi_{\k} - \sum_{\k'}U_{\k-\k'}\phi_{\k'},
    \label{eq:schrodinger equation DX} \\
    E_\text{IX}\psi_{\k} &= \bar\epsilon_{\k}\psi_{\k} - \sum_{\k'}U^d_{\k-\k'}\psi_{\k'}.
    \label{eq:schrodinger equation IX}
\end{align}
\end{subequations}
Note that these do not depend on the valley index.

In the following, we focus on the ground-state $1s$ solutions of Eqs.~\eqref{eq:schrodinger equation DX} and \eqref{eq:schrodinger equation IX}, and denote the corresponding momentum-space wave functions by $\Phi_{\k}$ and $\Psi_{\k}$, with eigenenergies $E^\text{DX}_{0}$ and $E^\text{IX}_{0}$. We choose the phase such that the wave functions are real. The energies are measured from the corresponding electron-hole continuum, see Fig.~\ref{fig:schematic}(b), and therefore the DX binding energy $\varepsilon_\text{X}$ is related to its ground-state energy via $\varepsilon_\text{X}=-E^\text{DX}_{0}$. We also define the corresponding effective exciton size $a_\text{X}=1/\sqrt{2\mu\varepsilon_\text{X}}$. Note that both $\varepsilon_\text{X}$ and $a_\text{X}$ depend on the dielectric environment and other physical properties of the MoS$_2$ homobilayer, as summarized in Table~\ref{tab:parameters}. We note that similar results for the exciton problem in a bilayer were recently obtained in Ref.~\cite{Maslova2024}.

In Fig.~\ref{fig:Scheq}, we show (a) the DX and IX eigenenergies as a function of environmental dielectric constant $\kappa$ at fixed interlayer separation $d$ and (b,c) the momentum-space wave functions of the DX and IX for vacuum and hBN encapsulation. As $\kappa$ increases, dielectric screening from the surrounding environment suppresses the long-range Coulomb attraction, thereby reducing both the DX and IX binding energies. The resulting weaker binding increases the real-space extent of the excitons, which is reflected in momentum space by wave functions that are more narrowly peaked around $\k=0$. We also observe that the IX is generally more weakly bound than the DX at the same $\kappa$, since the interlayer attraction is reduced compared to the intralayer attraction. Correspondingly, the IX wave function is more narrowly peaked in momentum space than the DX wave function.

\section{Microscopic description of exciton-exciton scattering}
\label{sec:microscopic description of exciton-exciton scattering}
Determining the exact interaction properties of two excitons requires solving the four-body \sch equation for two electrons and two holes described by the Hamiltonian in Eq.~\eqref{eq:electron-hole Hamiltonian}. However, the long-range nature of the Coulomb interactions and the presence of multiple breakup channels make the exact solution of the four-body Coulomb problem elusive. Instead, here we take advantage of the fact that an electron-hole pair can form a tightly bound exciton state, which allows us to reduce the four-body Coulomb problem to an effective two-body problem by projecting the four-particle Hilbert space onto the subspace spanned by two-exciton states. Such an effective two-exciton description provides a benchmark for microscopic treatments of exciton-exciton scattering~\cite{Ciuti1998,Tassone1999,Glazov2009}. In particular, in the following subsections we focus on the Born approximation for exciton-exciton scattering, adapted to a bilayer geometry~\cite{Byrnes2014,Nalitov2014a,Nalitov2014b,Maslova2024}. As we discuss in Secs.~\ref{sec:hybrid exciton} and \ref{sec:multichannel approach to scattering of hybrid excitons}, we can connect this microscopic description to an effective two-body pseudopotential of exciton-exciton interactions, which enables fully non-perturbative calculations of exciton-exciton scattering using a $T$-matrix approach.

\subsection{Exciton operators}
To implement the projection onto the two-exciton subspace, we introduce operators for an exciton as a bound electron-hole pair with an internal wave function. In a naturally stacked 2H MoS$_2$ homobilayer, it is convenient to label the two valley sectors shown in Fig.~\ref{fig:schematic}(b) by a single index $\eta \in \{1,2\}$, where $\eta=1$ denotes the $K'$ sector and $\eta=2$ denotes the $K$ sector. As above, we consider the electrons and holes to be spin polarized. For DXs, $\eta$ fixes the layer and valley of the constituent electron and hole, where $(l,\xi)=(1,K')$ for $\eta=1$ and $(l,\xi)=(2,K)$ for $\eta=2$. For IXs, $\eta$ labels the valley sector while simultaneously fixing the dipole orientation. Taking the $+z$ axis to point from the bottom layer ($l=1$) to the top layer ($l=2$), we choose $\eta=1$ to correspond to a dipole moment along $+z$ and $\eta=2$ to $-z$, as shown in Fig.~\ref{fig:schematic}(a).

For each $\eta$, the annihilation operator $\hat{x}_{\Q,\eta}$ corresponding to one of the two ground-state direct excitons shown in Fig.~\ref{fig:schematic}(b) takes the form
\begin{subequations}
    \label{eq:microscopic DX operator}
\begin{align}
    \hat{x}_{\Q,1} &= \sum_{\k} \Phi_{\k} \hat{e}_{\k+\gamma_{e}\Q,1K'}\hat{h}_{-\k+\gamma_{h}\Q,1K'}, \\
    \hat{x}_{\Q,2} &= \sum_{\k} \Phi_{\k}\hat{e}_{\k+\gamma_{e}\Q,2K}\hat{h}_{-\k+\gamma_{h}\Q,2K},
\end{align}
\end{subequations}
where $\Q$ is the total exciton momentum, $\k$ is the relative electron-hole momentum, and $\Phi_{\k}$ is the ground-state solution of Eq.~\eqref{eq:schrodinger equation DX} with normalization $\sum_\k |\Phi_{\k}|^2 = 1$. The coefficients $\gamma_\text{e} = m_\text{e}/m_\text{X}$ and $\gamma_\text{h} = m_\text{h}/m_\text{X}$ partition the total momentum $\Q$ between the electron and hole according to their masses, where $m_\text{X} = m_\text{e} + m_\text{h}$ corresponds to the exciton mass. Similarly, the ground-state IX annihilation operator $\hat{y}_{\Q,\eta}$ takes the form
\begin{subequations}
    \label{eq:microscopic IX operator}
\begin{align}
    \hat{y}_{\Q,1} &= \sum_{\k} \Psi_{\k} \hat{e}_{\k+\gamma_{e}\Q,1K'}\hat{h}_{-\k+\gamma_{h}\Q,2K'}, \\ 
    \hat{y}_{\Q,2} &= \sum_{\k} \Psi_{\k} \hat{e}_{\k+\gamma_{e}\Q,2K}\hat{h}_{-\k+\gamma_{h}\Q,1K},
\end{align}
\end{subequations}
where $\Psi_{\k}$ is the ground-state solution of Eq.~\eqref{eq:schrodinger equation IX} with normalization $\sum_\k |\Psi_{\k}|^2 = 1$. In the following, we use $\bar{\eta}$ to denote the index opposite to $\eta$.

\subsection{Born approximation}
To formulate the Born approximation for exciton-exciton scattering, we consider the two-exciton subspace at zero momentum, spanned by the states
\begin{equation}
    \ket{X_{\nu_1},X_{\nu_2}} = \hat{X}^{\dagger}_{\0,\nu_1}\hat{X}^{\dagger}_{\0,\nu_2}\ket{0}.
    \label{eq:general two-exciton state}
\end{equation}
Here, $\hat{X}^{\dagger}_{\0,\nu}$ creates an exciton with zero momentum, and the collective label $\nu$ specifies both the exciton species and the valley sector. The Born interaction constant is then obtained by projecting the electron-hole Hamiltonian in Eq.~\eqref{eq:electron-hole Hamiltonian} onto the two-exciton subspace, yielding~\cite{Levinsen2019a}
\begin{align}
    g^{\nu_1\nu_2}_\text{X-X} = 
     \frac{\bra{X_{\nu_1},X_{\nu_2}}
   \bigl(\hat{H}_{\text{eh}}-\mathcal{E}_{\nu_1}-\mathcal{E}_{\nu_2}\bigr)
    \ket{X_{\nu_1},X_{\nu_2}}}{1+\delta_{\nu_1\nu_2}}.
    \label{eq:general Born matrix element}
\end{align}
Here, $\mathcal{E}_{\nu_i}$ denotes the energy of an isolated zero-momentum exciton in the state $X_{\nu_i}$, as determined from the corresponding \sch equation in Eq.~\eqref{eq:full schrodinger equations}, with the appropriate electron-hole continuum threshold retained. Its relative-motion energy $E_{\nu_i}$ is obtained from the relative-motion \sch equation in Eq.~\eqref{eq:relative schrodinger equations}. The subtraction of $\mathcal{E}_{\nu_1}+\mathcal{E}_{\nu_2}$ removes the energies of the two isolated excitons, leaving only the interaction-induced contribution to the two-exciton matrix element. The factor $1+\delta_{\nu_1\nu_2}$ accounts for the symmetry factor associated with two identical excitons.

The Born approximation is therefore determined entirely by the interexciton Coulomb interactions, i.e., the Coulomb interactions between charge carriers belonging to different excitons. 
Evaluating these terms between two-exciton states expressed in terms of electron and hole operators yields both direct (Hartree) and exchange contributions. The direct term describes Coulomb scattering without exchanging charges between the two excitons, whereas the exchange term arises from the fermionic indistinguishability of electrons and/or holes in the two-exciton state. Consequently, if the two excitons contain only distinguishable charge carriers, the Born approximation reduces to the direct contribution. By contrast, when the two excitons share indistinguishable electrons and/or holes, one must additionally include the corresponding exchange processes. In the following, we apply this general structure to the intravalley and intervalley scattering channels relevant to bilayer MoS$_2$.

\begin{figure}[tp]
    \begin{minipage}{0.48\textwidth}
    \centering
    \includegraphics[width=\textwidth]{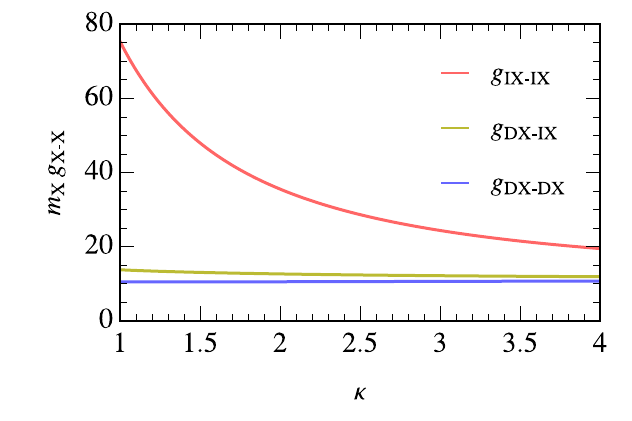}
    \end{minipage}
    \caption{Born approximation for intravalley exciton-exciton scattering at zero momentum as a function of dielectric constant, where for brevity, we suppress the index $\eta$. The red, yellow, and blue lines correspond to intravalley DX-DX, DX-IX, and IX-IX scattering, respectively. All parameters are listed in Table~\ref{tab:parameters}.}
    \label{fig:gXX}
\end{figure}

\subsection{Intravalley exciton-exciton scattering}
\label{subsec:intravalley exciton-exciton scattering}
We first consider intravalley scattering, where the two excitons belong to the same valley sector, e.g., $K'$ in Fig.~\ref{fig:schematic}(b). In this case, the two excitons share at least one indistinguishable constituent, so exchange processes contribute in addition to any surviving direct term. For intravalley DX-DX scattering, the two DXs occupy the same layer and valley. Therefore, both electrons and holes are indistinguishable, and the Born approximation contains exchange contributions from both types of charge carriers. Following Ref.~\cite{Levinsen2019a}, we obtain
\begin{align}
    &g^{\eta\eta}_\text{DX-DX} = \frac{1}{2}\bra{0}\hat{x}_{\0,\eta}\hat{x}_{\0,\eta}(\hat{H}_\text{eh}-2\mathcal{E}^\text{DX}_{0,\eta})\hat{x}^{\dagger}_{\0,\eta}\hat{x}^{\dagger}_{\0,\eta}\ket{0} \notag \\ 
    &\quad = 2\sum_{\k}({\bar\epsilon}_{\k}-E^\text{DX}_{0})\Phi^{4}_{\k} - 2\sum_{\k\k'}U_{\k-\k'}\Phi^{2}_{\k}\Phi^{2}_{\k'},
    \label{eq:intravalley DX-DX scattering}
\end{align}
where the prefactor $1/2$ accounts for the symmetry factor of two identical excitons. The direct term vanishes in the zero-momentum limit due to the charge neutrality of the DXs within a single layer, leaving only exchange contributions from electrons and holes. The expression in Eq.~\eqref{eq:intravalley DX-DX scattering} is equivalent to the corresponding Born approximation derived in Ref.~\cite{Pico2025} in the context of a TMD monolayer, and it reproduces the Coulomb limit in Refs.~\cite{Ciuti1998,Tassone1999}.

For intravalley (parallel) IX-IX scattering, the electrons and holes remain indistinguishable within their respective layers, and exchange processes contribute to the Born approximation. In contrast to the DX-DX case, however, the direct term does not vanish in the zero-momentum limit because each IX carries a static out-of-plane dipole moment across the two layers. Specifically, we find
\begin{align}
    &g^{\eta\eta}_\text{IX-IX} = \frac{1}{2}\bra{0}\hat{y}_{\0,\eta}\hat{y}_{\0,\eta}(\hat{H}_\text{eh}-2\mathcal{E}^\text{IX}_{0,\eta})\hat{y}^{\dagger}_{\0,\eta}\hat{y}^{\dagger}_{\0,\eta}\ket{0} \notag \\ 
    &\quad = \frac{4\pi e^{2}d}{\kappa} + 2\sum_{\k}({\bar\epsilon}_{\k}-E^\text{IX}_{0})\Psi^{4}_{\k} - 2\sum_{\k\k'}U_{\k-\k'}\Psi^{2}_{\k}\Psi^{2}_{\k'}.
    \label{eq:intravalley IX-IX scattering}
\end{align}
Here, the first term is the direct contribution obtained from the difference between the intra- and interlayer Coulomb interactions in the zero-momentum limit
\begin{equation}
    \lim_{\q\to\0} \left( U_{\q} + U_{\q} - U^{d}_{\q} - U^{d}_{\q} \right) = \frac{4\pi e^{2}d}{\kappa},
\end{equation}
where the two positive terms correspond to the electron-electron and hole-hole repulsion, while the two negative terms arise from the electron-hole attraction between different excitons. The remaining terms in Eq.~\eqref{eq:intravalley IX-IX scattering} are exchange contributions associated with indistinguishable electrons and holes, and hence the functional form is similar to Eq.~\eqref{eq:intravalley DX-DX scattering} above.

Finally, for intravalley DX-IX scattering, only the electrons are indistinguishable between the two excitons, as shown in Fig.~\ref{fig:schematic}(b), and thus the interaction is dominated by electron-exchange processes. We therefore again find a natural extension of Eq.~\eqref{eq:intravalley DX-DX scattering}:
\begin{align}
    g^{\eta\eta}_\text{DX-IX} &= \bra{0}\hat{y}_{\0,\eta}\hat{x}_{\0,\eta}(\hat{H}_\text{eh}-\mathcal{E}^\text{DX}_{0,\eta}-\mathcal{E}^\text{IX}_{0,\eta})\hat{x}^{\dagger}_{\0,\eta}\hat{y}^{\dagger}_{\0,\eta}\ket{0} \notag \\ 
    &= \sum_{\k}(2{\bar\epsilon}_{\k}-E^\text{DX}_{0}-E^\text{IX}_{0})\Phi_{\k}^{2}\Psi^{2}_{\k} \notag \\ 
    &\quad - \sum_{\k\k'}(U_{\k-\k'}+U^{d}_{\k-\k'})\Phi^{2}_{\k}\Psi^{2}_{\k'}.
    \label{eq:intravalley DX-IX scattering}
\end{align}
The direct term cancels in the zero momentum limit because the four interexciton Coulomb terms sum to zero.

In Fig.~\ref{fig:gXX}, we show the interaction constants for intravalley DX-DX, IX-IX, and DX-IX scattering obtained from Eqs.~\eqref{eq:intravalley DX-DX scattering}, \eqref{eq:intravalley IX-IX scattering}, and \eqref{eq:intravalley DX-IX scattering} as functions of environmental dielectric constant $\kappa$. We see that all the interactions are repulsive and we clearly observe a pronounced enhancement of the IX-IX scattering as $\kappa\to 1$, consistent with the results of Ref.~\cite{Maslova2024}. This enhancement is primarily driven by the direct term in the first term of Eq.~\eqref{eq:intravalley IX-IX scattering}, which reflects the dipole-dipole repulsion between parallel IXs and increases as the environmental screening is reduced. At the same time, decreasing $\kappa$ also increases the exciton binding energies and modifies the internal wave functions, which in turn affects the exchange contributions through the exciton wave function overlap. For DX-DX scattering, the exchange contributions vary only weakly with $\kappa$ and exhibit a slight decrease towards the vacuum limit, in agreement with the monolayer case~\cite{Pico2025}. On the other hand, the DX-IX scattering is modestly enhanced as $\kappa$ decreases, but remains small compared with the IX-IX scattering.

\subsection{Intervalley exciton-exciton scattering}
\label{subsec:intervalley exciton-exciton scattering}
We now consider intervalley scattering, for which the two excitons belong to opposite valley sectors, i.e., $K$ and $K'$ in Fig.~\ref{fig:schematic}(b). In this case, the electrons and holes associated with the two excitons are fully distinguishable, and the Born approximation is determined solely by the direct contribution. For intervalley DX-DX scattering, the Born approximation reduces to
\begin{align}
    g^{\eta\bar{\eta}}_\text{DX-DX} &= \bra{0}\hat{x}_{\0,\bar{\eta}}\hat{x}_{\0,\eta}(\hat{H}_\text{eh}-\mathcal{E}^\text{DX}_{0,\eta}-\mathcal{E}^\text{DX}_{0,\bar{\eta}})\hat{x}^{\dagger}_{\0,\eta}\hat{x}^{\dagger}_{\0,\bar{\eta}}\ket{0} \notag \\
    &= \lim_{\q\to\0} \left[ U^{d}_{\q} + U^{d}_{\q} - U^{d}_{\q} - U^{d}_{\q} \right] \notag \\
    &= 0.
    \label{eq:intervalley DX-DX scattering}
\end{align}
All four interexciton Coulomb interactions are given by the interlayer Keldysh potential in Eq.~\eqref{eq:interlayer Keldysh potential}, which cancel exactly in the zero-momentum limit. Thus, the direct contribution between two intervalley DXs vanishes.

We next consider intervalley IX-IX scattering, corresponding to two IXs with antiparallel dipole moments. In this case, the Born approximation takes the form
\begin{align}
    g^{\eta\bar{\eta}}_\text{IX-IX} &= \bra{0}\hat{y}_{\0,\bar{\eta}}\hat{y}_{\0,\eta}(\hat{H}_\text{eh}-\mathcal{E}^\text{IX}_{0,\eta}-\mathcal{E}^\text{IX}_{0,\bar{\eta}})\hat{y}^{\dagger}_{\0,\eta}\hat{y}^{\dagger}_{\0,\bar{\eta}}\ket{0} \notag \\ 
    &= \lim_{\q\to\0} \left[ U^{d}_{\q} + U^{d}_{\q} - U_{\q} - U_{\q} \right] \notag \\
    &= -\frac{4\pi e^{2}d}{\kappa}.
    \label{eq:intervalley IX-IX scattering}
\end{align}
The two positive terms correspond to the interlayer electron-electron and hole-hole repulsion, whereas the two negative terms correspond to the intralayer electron-hole attraction between different excitons. Their difference yields a finite negative interaction, reflecting the attractive dipole-dipole interaction between antiparallel IXs.

Similarly, the Born approximation for DX-IX scattering is
\begin{align}
    g^{\eta\bar{\eta}}_\text{DX-IX} &= \bra{0}\hat{y}_{\0,\bar{\eta}}\hat{x}_{\0,\eta}(\hat{H}_\text{eh}-\mathcal{E}^\text{DX}_{0,\eta}-\mathcal{E}^\text{IX}_{0,\bar{\eta}})\hat{x}^{\dagger}_{\0,\eta}\hat{y}^{\dagger}_{\0,\bar{\eta}}\ket{0} \notag \\ 
    &= \lim_{\q\to\0} \left[ U_{\q} + U^{d}_{\q} - U_{\q} - U^{d}_{\q} \right] \notag \\
    &= 0.
    \label{eq:intervalley DX-IX scattering}
\end{align}
Here, the interexciton Coulomb interactions involve both intra- and interlayer Keldysh potentials, but the repulsive and attractive contributions cancel exactly in the zero-momentum limit.

Having a vanishing zero-momentum Born approximation does not necessarily imply that the interaction between two excitons is weak. A prominent example is that of two direct excitons with opposite electronic spins in a monolayer. In this case, the exchange contribution to scattering is absent because the constituent charge carriers are distinguishable, and the direct contribution vanishes in the zero-momentum limit due to overall charge neutrality~\cite{Ciuti1998,Tassone1999}. Nevertheless, such excitons can still form a biexciton bound state, which is a clear signature of strong effective exciton-exciton interactions. Indeed, biexcitons have been observed both in GaAs quantum wells~\cite{Kavokin2017microcavities} and in TMD monolayers~\cite{Mai2014,You2015,Hao2017,Nagler2018}.

Similarly, we may ask whether an intervalley DX-DX bound state is possible. However, given that these direct excitons occupy different layers, which strongly suppresses the short-range overlap needed for biexciton formation, we consider such a bound state highly unlikely. This is consistent with previous studies of bilayer excitons, which show that biexcitons only exist for a very small layer separation~\cite{Schindler2008,Lee2009}, well below the regime relevant to homobilayer MoS$_2$. Therefore, in agreement with the Born approximation, it is likely that intervalley DX-DX scattering is small.

The other vanishing Born approximation is that of intervalley DX-IX scattering. This case is highly non-trivial, as the direct exciton, together with the same-layer hole of the IX, may form an intralayer positively charged trion bound state, analogous to the tightly bound trions observed in monolayer MoS$_2$~\cite{Mak2013}. The remaining electron in the opposite layer may then bind to this trion through the long-range Coulomb interaction, raising the possibility of a four-body bound state. Whether such a state actually exists in the present bilayer geometry, however, remains an open question. 

Finally, Eq.~\eqref{eq:intervalley IX-IX scattering} implies that the intervalley IX-IX scattering is attractive, as anticipated for antiparallel dipoles. However, this does not necessarily imply the existence of an intervalley IX-IX bound state. Instead, the two intervalley IXs can reconfigure their constituent charges into two (optically dark) intralayer excitons in different layers, which can lie lower in energy than the original IX pair. By the same argument as above~\cite{Schindler2008,Lee2009}, these spatially separated excitons are unlikely to form bound states. The reconfiguration of charges is therefore likely to dominate the intervalley IX-IX scattering, but its precise description is beyond the scope of the present work.

The above arguments demonstrate that intervalley exciton-exciton scattering currently features several important open questions. Therefore, in the following we focus on exciton-exciton interactions in the better-understood intravalley case. Here, there are no bound states and the interactions are expected to be repulsive, in agreement with the Born approximation above.

\section{Hybrid exciton}
\label{sec:hybrid exciton}
We now formulate an effective excitonic description of the hybrid exciton, i.e., an electrically tunable superposition of a DX and an IX in a naturally stacked 2H MoS$_2$ homobilayer. The parameters of our excitonic model are derived from the underlying electronic Hamiltonian in Eq.~\eqref{eq:electron-hole Hamiltonian}.

\subsection{Coupled oscillator model}
\label{subsec:coupled oscillator model}
The hybridization between DX and IX states originates from interlayer hole tunneling. Microscopically, this process is described by the additional term
\begin{align}
    \hat{H}_{t_\text{h}} = \frac{t_\text{h}}{2} \sum_{\k,\xi} \left( \hat{h}^{\dagger}_{\k,1\xi}\hat{h}_{\k,2\xi} + \hat{h}^{\dagger}_{\k,2\xi}\hat{h}_{\k,1\xi} \right),
    \label{eq:microscopic hole tunneling}
\end{align}
where $t_\text{h}$ denotes the microscopic hole-tunneling parameter. The corresponding microscopic electron-hole problem can be written as a coupled \sch equation for the DX and IX components of the exciton wave function
\begin{align}
    E^{(\eta)}\phi_{\k} &= (\bar\epsilon_{\k} - E^\text{DX}_{0})\phi_{\k} - \sum_{\k'}U_{\k-\k'}\phi_{\k'} + \frac{t_\text{h}}{2}\psi_{\k}, \notag \\
    E^{(\eta)}\psi_{\k} &= (\bar\epsilon_{\k} + \delta_\text{IX} \mp \Delta - E^\text{IX}_{0})\psi_{\k} \notag \\ 
    &\hspace{5mm} - \sum_{\k'}U^d_{\k-\k'}\psi_{\k'} + \frac{t_\text{h}}{2}\phi_{\k}.
    \label{eq:coupled schrodinger equation}
\end{align}
Here, the energy $E^{(\eta)}$ is measured relative to the uncoupled DX ground-state energy. Using the relations between the exciton energies and their corresponding continuum thresholds in Eq.~\eqref{eq:relative exciton energies}, the ground-state DX and IX energies are $\mathcal{E}^\text{DX}_{0,l\xi} = \delta^\text{h}_{l\xi} + E^\text{DX}_{0}$ and $\mathcal{E}^\text{IX}_{0,l\xi} = \delta^\text{h}_{\bar{l}\xi} + E^\text{IX}_{0}$, respectively. We therefore define the bare IX-DX detuning at zero electric field as
\begin{equation}
    \delta_\text{IX} \equiv \mathcal{E}^\text{IX}_{0,l\xi} - \mathcal{E}_{0,l\xi}^\text{DX},
\end{equation}
which is the energy difference measured spectroscopically. Thus, $\delta_\text{IX}$ is the energy of the uncoupled IX ground state relative to the uncoupled DX ground state at zero electric field. We additionally incorporate the effect of an out-of-plane electric field through the Stark shift $\mp\Delta$ of the two oppositely oriented IX configurations, where $\Delta$ is proportional to the applied electric field~\cite{Lorchat2021}. The sign in front of $\Delta$ is determined by the dipole orientation and is taken to be $-$ for $\eta=1$ and $+$ for $\eta=2$.

The coupled \sch equation in Eq.~\eqref{eq:coupled schrodinger equation} can be approximated as a coupled oscillator model by expanding the DX and IX components in the eigenstates of the uncoupled DX and IX \sch equations. Denoting the corresponding exciton creation operators by $\hat{x}^{\dagger}_{m,\Q,\eta}$ and $\hat{y}^{\dagger}_{n,\Q,\eta}$, where $m$ and $n$ label the internal DX and IX states, the coupled-oscillator excitonic Hamiltonian takes the form
\begin{align}
    \hat{H}^{(\eta)}_\text{CO} &= \sum_{m,\Q} \epsilon^\text{DX}_{m,\Q}\hat{x}^{\dagger}_{m,\Q,\eta}\hat{x}_{m,\Q,\eta} + \sum_{n,\Q}
    \epsilon^\text{IX}_{n,\Q,\eta}\hat{y}^{\dagger}_{n,\Q,\eta}\hat{y}_{n,\Q,\eta} \notag \\
    &+ \frac{1}{2}
    \sum_{m,n}\sum_{\Q} t_{mn} \left( \hat{x}^{\dagger}_{m,\Q,\eta}\hat{y}_{n,\Q,\eta} + \hat{y}^{\dagger}_{n,\Q,\eta} \hat{x}_{m,\Q,\eta} \right).
    \label{eq:coupled oscillator model}
\end{align}
Here, the dispersions of the uncoupled DX and IX modes are defined as
\begin{align}
    \epsilon^\text{DX}_{m,\Q} &= \frac{Q^2}{2m_\text{DX}} + \left( E^\text{DX}_{m}-E^\text{DX}_{0} \right), \\ 
    \epsilon^\text{IX}_{n,\Q,\eta} &= \frac{Q^2}{2m_\text{IX}} + \delta_\text{IX} \mp \Delta + \left(E^\text{IX}_{n}-E^\text{IX}_{0}\right),
\end{align}
where $E^\text{DX}_{m}$ and $E^\text{IX}_{n}$ are the internal eigenenergies of the uncoupled DX and IX \sch equations, respectively. The indices $m$ and $n$ label the corresponding internal exciton states, ordered by energy, with $m=n=0$ denoting the ground-state $1s$ DX and IX. Thus, we have $\epsilon^\text{DX}_{0,\Q}=Q^2/2m_\text{DX}$ and $\epsilon^\text{IX}_{0,\Q,\eta}=Q^2/2m_\text{IX}+\delta_\text{IX}\mp\Delta$.

The tunneling matrix elements are obtained by treating the microscopic hole-tunneling term in Eq.~\eqref{eq:microscopic hole tunneling} to first order and projecting it onto the corresponding uncoupled DX and IX eigenstates. This perturbative approach is valid as long as $t_\text{h}$ is smaller than the DX and IX binding energies, and gives
\begin{equation}
    \frac{t_{mn}}{2} = \bra{0}\hat{x}_{m,\Q,\eta}\hat{H}_{t_\text{h}}
    \hat{y}^{\dagger}_{n,\Q,\eta}\ket{0}.
\end{equation}
Evaluating this matrix element using the microscopic exciton operators gives
\begin{equation}
    t_{mn} = t_\text{h} \sum_{\k} \Phi^{\ast}_{m,\k}
    \Psi_{n,\k},
    \label{eq:tunneling matrix element}
\end{equation}
where $\Phi_{m,\k}$ and $\Psi_{n,\k}$ are the $m$-th DX and $n$-th IX wave functions, respectively, and we choose the phase convention such that the wave functions are real. Thus, the coupling strength between DX and IX states in the excitonic Hamiltonian is given by the microscopic hole tunneling weighted by the overlap between the corresponding two exciton wave functions. 

\begin{figure}[tp]
    \begin{minipage}{0.48\textwidth}
    \centering
    \includegraphics[width=\textwidth]{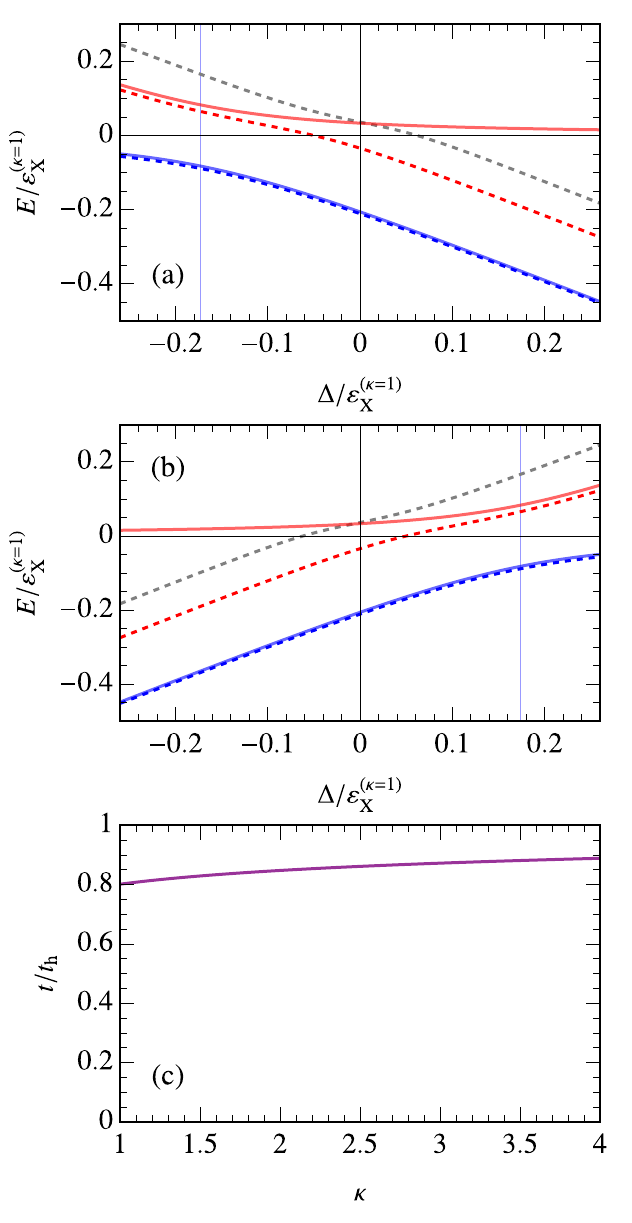}
    \end{minipage}
    \caption{(a,b) Hybrid-exciton energies as a function of the Stark shift $\Delta$ at zero momentum with $\kappa=1$ for (a) $\eta=1$ and (b) $\eta=2$. We compare the lowest three eigenvalues obtained from the microscopic coupled \sch equation (blue, red, and gray dashed lines) with the two hybridized modes obtained from the two-mode coupled-oscillator model (blue and red solid lines). The coupled-oscillator model uses $t=76$ meV, while the microscopic calculation uses the corresponding $t_\text{h}$ determined from Eq.~\eqref{eq:tfromth}. The blue vertical lines indicate the resonance condition between the $1s$ IX and $1s$ DX modes, $\delta_\text{IX}\mp\Delta=0$, for $\delta_\text{IX}/\varepsilon^{(\kappa=1)}_\text{X} = -0.17$. The minus sign applies to $\eta=1$ in (a), while the plus sign applies to $\eta=2$ in (b). (c) Overlap between the ground-state DX and IX wave functions as a function of $\kappa$. In (a,b), we assume $m_\text{DX} = m_\text{IX}\equiv m_\text{X}$, with $m_\text{X}=m_\text{e}+m_\text{h}$. The remaining parameters are listed in Table~\ref{tab:parameters}.}
    \label{fig:coupled oscillator}
\end{figure}

The commonly applied theoretical description of the hybrid exciton is the two-mode coupled-oscillator model, in which only the ground-state DX and IX modes with $m=n=0$ are retained in Eq.~\eqref{eq:coupled oscillator model}. To assess the validity of this approximation, Figs.~\ref{fig:coupled oscillator}(a,b) compare the hybrid-exciton energies obtained from the microscopic coupled \sch equation in Eq.~\eqref{eq:coupled schrodinger equation} with those obtained from the two-mode coupled-oscillator model as a function of the Stark shift $\Delta$. The two descriptions agree well when the $1s$ IX mode is near resonance with the $1s$ DX mode, $\delta_\text{IX}\mp\Delta\simeq0$, where the relevant hybrid-exciton branches are dominated by these two states. Away from this regime, the full microscopic calculation can contain additional modes associated with excited exciton states, which primarily affect the upper hybridized branch and lead to deviations from the two-mode description. We note, however, that such modes are not observed as resolved resonances in photoluminescence or reflectance in experiment, likely because they have weak oscillator strength or are strongly broadened. We therefore use the two-mode coupled-oscillator model in the following as an effective description of the optically relevant hybrid exciton modes, focusing on the regime near the resonance between the $1s$ DX and $1s$ IX modes for $\Delta\geq0$.

The effective tunneling rate entering the two-mode coupled-oscillator model is given by Eq.~\eqref{eq:tunneling matrix element} with $m=n=0$. This leads to
\begin{equation}
    t \equiv t_{00} = t_\text{h} \sum_{\k}\Phi^{\ast}_{\k}\Psi_{\k}.
    \label{eq:tfromth}
\end{equation}
In practice, experiments typically measure the effective coupling $t$ through the DX-IX anticrossing, rather than the bare single-particle tunneling amplitude $t_\text{h}$. We thus use $t$ directly as a phenomenological parameter in the excitonic Hamiltonian. In principle, Eq.~\eqref{eq:tfromth} implies a weak dependence of $t$ on the dielectric constant $\kappa$ through the DX and IX wave functions. However, within the parameter regime explored in this work, we find that the corresponding overlap varies by only about $10\%$, as shown in Fig.~\ref{fig:coupled oscillator}(c). We therefore fix $t$ to a constant value of $t=76$\,meV, which lies within the range of values reported in Refs.~\cite{Deilmann2018,Leisgang2020,Louca2023}.

\subsection{Excitonic model}
\label{subsec:excitonic model}
In the following, we consider hybrid excitons within a fixed sector $\eta$. This reflects the fact that the DX-IX hybridization occurs independently within each $\eta$ sector and that we focus on the corresponding intravalley scattering problem, for which it is sufficient to retain only effective exciton-exciton interactions acting within the same $\eta$ sector. Scattering between hybrid excitons belonging to different valleys would instead constitute a distinct intervalley problem, which could involve additional correlated states such as biexcitons, and is therefore beyond the scope of the present work.

We thus consider a model of hybrid excitons as superpositions of the ground-state DX and IX within a given valley sector $\eta$. This is obtained by restricting the multimode coupled-oscillator Hamiltonian in Eq.~\eqref{eq:coupled oscillator model} to the $m=n=0$ sector. The resulting Hamiltonian is $\hat{H}^{(\eta)} = \hat{H}^{(\eta)}_{0} + \hat{V}^{(\eta)}$, with the non interacting part
\begin{align}
    \hat{H}^{(\eta)}_{0} = &\sum_{\Q} \Big[ \epsilon^{\text{DX}}_{\Q}\hat{x}^{\dagger}_{\Q,\eta}\hat{x}_{\Q,\eta} + \epsilon^{\text{IX}}_{\Q,\eta}\hat{y}^{\dagger}_{\Q,\eta}\hat{y}_{\Q,\eta} \notag \\ 
    &\quad + \frac{t}{2}  \left(\hat{x}^{\dagger}_{\Q,\eta}\hat{y}_{\Q,\eta}+\hat{y}^{\dagger}_{\Q,\eta}\hat{x}_{\Q,\eta}\right) \Big].
    \label{eq:non-interacting Hamiltonian hX}
\end{align}
The direct and indirect excitons are treated as structureless bosons that approximately correspond to the operators defined in Eqs.~\eqref{eq:microscopic DX operator} and \eqref{eq:microscopic IX operator}, with the corresponding DX and IX dispersions $\epsilon^{\text{DX}}_{\Q}=Q^2/2m_\text{DX}$ and $\epsilon^{\text{IX}}_{\Q,\eta}=Q^2/2m_\text{IX}+\delta_\text{IX}\mp\Delta$, respectively. The IX dispersion is parameterized by the bare IX-DX detuning $\delta_\text{IX}$ and the Stark shift $\Delta$ arising from the coupling of the indirect exciton to an applied out-of-plane electric field~\cite{Leisgang2020,Lorchat2021}. The sign in front of $\Delta$ encodes the two dipole orientations and is taken to be $-$ for $\eta=1$ and $+$ for $\eta=2$. The hybridization between the DX and IX arises from interlayer tunneling of the hole within the same valley, parameterized by the effective tunneling rate $t$, as discussed above.

Diagonalizing the non-interacting Hamiltonian in Eq.~\eqref{eq:non-interacting Hamiltonian hX} yields the two hybridized exciton modes
\begin{align}
    \hat{H}^{(\eta)}_{0} &= \sum_{\Q} \left[ E^{-}_{\Q,\eta}\hat{L}^{\dagger}_{\Q,\eta}\hat{L}_{\Q,\eta} + E^{+}_{\Q,\eta}\hat{U}^{\dagger}_{\Q,\eta}\hat{U}_{\Q,\eta} \right],
\end{align}
where $\hat{L}_{\Q,\eta}$ and $\hat{U}_{\Q,\eta}$, respectively, correspond to the annihilation operators of the lower (hX$^-$) and upper (hX$^+$) hybrid excitons. The hybrid exciton dispersions read
\begin{equation}
    E^{\pm}_{\Q,\eta} = \frac{1}{2}\left( \epsilon^{\text{IX}}_{\Q,\eta} + \epsilon^{\text{DX}}_{\Q} \pm \sqrt{(\epsilon^{\text{IX}}_{\Q,\eta}-\epsilon^{\text{DX}}_{\Q})^{2} + t^{2}} \right).
    \label{eq:hybrid exciton dispersions}
\end{equation}
The hybrid exciton operators are related to the DX and IX operators via the linear transformation
\begin{equation}
    \begin{pmatrix}
    \hat{L}_{\Q,\eta} \\[0.2cm] \hat{U}_{\Q,\eta}
    \end{pmatrix}
    = 
    \begin{pmatrix}
    X^{-}_{\Q,\eta} & Y^{-}_{\Q,\eta} \\[0.2cm]
    X^{+}_{\Q,\eta} & Y^{+}_{\Q,\eta} \\
    \end{pmatrix}
    \begin{pmatrix}
    \hat{x}_{\Q,\eta} \\[0.2cm] \hat{y}_{\Q,\eta}
    \end{pmatrix},
    \label{eq:hybrid exciton operators}
\end{equation}
where the transformation coefficients take the analytic forms $X^{-}_{\Q,\eta} = Y^{+}_{\Q,\eta} = u_{\Q,\eta}$ and $Y^{-}_{\Q,\eta} = -X^{+}_{\Q,\eta} = v_{\Q,\eta}$ with
\begin{align}
    u^{2}_{\Q,\eta} &= \frac{1}{2} \left( 1 + \frac{\epsilon^{\text{IX}}_{\Q,\eta}-\epsilon^{\text{DX}}_{\Q}}{E^{+}_{\Q,\eta}-E^{-}_{\Q,\eta}} \right), \notag \\[0.2cm] 
    v^{2}_{\Q,\eta} &= \frac{1}{2} \left( 1 - \frac{\epsilon^{\text{IX}}_{\Q,\eta}-\epsilon^{\text{DX}}_{\Q}}{E^{+}_{\Q,\eta}-E^{-}_{\Q,\eta}} \right),
    \label{eq:transformation coefficients hX}
\end{align}
and satisfy $u^{2}_{\Q,\eta}+v^{2}_{\Q,\eta} = 1$. Physically, the squared coefficients give the DX and IX weights of the hybrid exciton modes. For the hX$^-$ mode, the DX and IX fractions are $u_{\Q,\eta}^{2}$ and $v_{\Q,\eta}^{2}$, respectively, whereas for the hX$^+$ mode, the DX and IX fractions are $v_{\Q,\eta}^{2}$ and $u_{\Q,\eta}^{2}$.

In a 2H MoS$_2$ homobilayer, the electron and hole effective masses are, to a good approximation, independent of layer and valley, as we have implicitly assumed in the microscopic Hamiltonian in Eq.~\eqref{eq:electron-hole Hamiltonian}. Therefore, we set $m_\text{DX} = m_\text{IX} \equiv m_\text{X}$ with the exciton mass $m_\text{X} = m_\text{e} + m_\text{h}$. In this case, the transformation coefficients in Eq.~\eqref{eq:transformation coefficients hX} become momentum independent.

\subsection{Interaction potentials}
\label{subsec:interaction potentials}
Interactions between hybrid excitons originate from their DX and IX constituents. Within the excitonic model, we describe these processes by the potential operator
\begin{align}
    \hat{V}^{(\eta)} &= \frac{1}{2}\sum_{\Q\Q'\q} V_{\text{DX-DX}}(\q) \,\hat{x}^{\dagger}_{\Q+\q,\eta}\hat{x}^{\dagger}_{\Q'-\q,\eta}\hat{x}_{\Q',\eta}\hat{x}_{\Q,\eta} \notag \\
    &+ \frac{1}{2}\sum_{\Q\Q'\q} V_{\text{IX-IX}}(\q) \,\hat{y}^{\dagger}_{\Q+\q,\eta}\hat{y}^{\dagger}_{\Q'-\q,\eta}\hat{y}_{\Q',\eta}\hat{y}_{\Q,\eta} \notag \\
    &+ \sum_{\Q\Q'\q} V_{\text{DX-IX}}(\q) \,\hat{x}^{\dagger}_{\Q+\q,\eta}\hat{y}^{\dagger}_{\Q'-\q,\eta}\hat{y}_{\Q',\eta}\hat{x}_{\Q,\eta} \, ,
    \label{eq:potential operator}
\end{align}
where again the factor $1/2$ originates from indistinguishability.
Here, $V_\text{X-X}(\q)$ is the 2D Fourier transform of a real-space potential
\begin{equation}
    V_\text{X-X}(\q)=\int d^2\mathbf{r}\, e^{-i\q\cdot\mathbf{r}}\,V_\text{X-X}(\mathbf{r}).
\end{equation}

We choose the form of the effective potentials $V_\text{X-X}(\q)$ for DX-DX, IX-IX, and DX-IX interactions such that they are consistent with the underlying microscopic description of intravalley exciton-exciton scattering in Sec.~\ref{subsec:intravalley exciton-exciton scattering}. For the DX-DX and DX-IX interaction potentials, the microscopic Born approximation yields short-range exchange-dominated scattering processes in the intravalley sector [Eqs.~\eqref{eq:intravalley DX-DX scattering} and \eqref{eq:intravalley DX-IX scattering}]. We therefore model these interactions by short-range soft-core pseudopotentials with range set by the effective exciton size $a_\text{X}$ determined from the DX \sch equation in Eq.~\eqref{eq:schrodinger equation DX}. In real space, these take the forms
\begin{align}
    V_\text{DX-DX}(\mathbf{r}) &= V^{\text{DX-DX}}_{0}\theta\left(a_\text{X}-r\right), \label{eq:VDX-DX} \\[0.2cm] 
    V_\text{DX-IX}(\mathbf{r}) &= V^{\text{DX-IX}}_{0}\theta\left(a_\text{X}-r\right), \label{eq:VDX-IX}
\end{align}
where $\theta(x)$ denotes the Heaviside step function. On the other hand, for IX-IX scattering, the interactions contain a long-range dipole-dipole repulsion for out-of-plane dipoles $V(\mathbf{r})\sim D^2/r^3$, where we define $D^2 = e^{2}d^{2}/\kappa$. At distances shorter than a cutoff $r_0$, the point-dipole form breaks down due to the composite nature of the excitons and exchange effects. We therefore model the regularized dipolar potential using a soft-core potential
\begin{equation}
    V_\text{IX-IX}(\mathbf{r}) = \frac{D^2}{r_{0}^{3}}\theta\left(r_{0}-r\right) + \frac{D^2}{r^{3}}\theta\left(r-r_{0}\right).
    \label{eq:VIX-IX}
\end{equation}

We emphasize that the Born approximation is conceptually useful because, in the absence of a lower-lying two-body (biexciton) bound state, it provides an upper bound on the interaction constant~\cite{Li2021a}. 
We therefore fix the pseudopotential parameters by requiring that the zero-momentum approximation reproduces the corresponding microscopic Born interaction constants
\begin{equation}
    g^{\eta\eta}_\text{X-X} = V_\text{X-X}(\q=0)
    =\int d^2\mathbf{r}\,V_\text{X-X}(\mathbf{r}).
\end{equation}
For the soft-core potentials in Eqs.~\eqref{eq:VDX-DX} and \eqref{eq:VDX-IX}, this procedure gives
\begin{align}
    V^{\text{DX-DX}}_{0} &= \frac{g^{\eta\eta}_\text{DX-DX}}{\pi a^{2}_\text{X}},
    \quad
    V^{\text{DX-IX}}_{0} = \frac{g^{\eta\eta}_\text{DX-IX}}{\pi a^{2}_\text{X}}.
\end{align}
For the regularized dipolar potential in Eq.~\eqref{eq:VIX-IX}, one finds
\begin{equation}
    r_{0} = \frac{3\pi D^{2}}{g^{\eta\eta}_\text{IX-IX}}.
\end{equation}
This procedure ensures that the upper bound for exciton-exciton interactions is the same within the microscopic approach in Sec.~\ref{sec:microscopic description of exciton-exciton scattering} and the multichannel approach presented in Sec.~\ref{sec:multichannel approach to scattering of hybrid excitons} below. 

\begin{figure*}[tp]
    \begin{minipage}{0.8\textwidth}
    \centering
    \includegraphics[width=\textwidth]{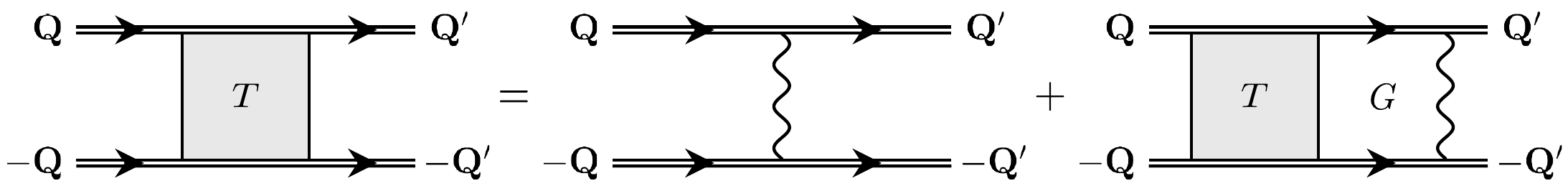}
    \end{minipage}
    \caption{Diagrammatic representation of the coupled $T$-matrix equation in Eq.~\eqref{eq:partial wave scattering integral equation}. Double lines denote hybridized excitons, with external momenta $\Q$ and $\Q'$ for the incoming and outgoing two-particle states, respectively. The vertical wavy line represents the effective exciton-exciton interaction, while the shaded square denotes the coupled $T$ matrix. The first diagram corresponds to the Born term. The second diagram represents repeated scattering through an intermediate two-particle propagator, together with a sum over intermediate channels $n$.}
    \label{fig:T-matrix equation}
\end{figure*}

\section{Multichannel approach to scattering of hybrid excitons}
\label{sec:multichannel approach to scattering of hybrid excitons}
Having defined the hybrid excitons and their effective interaction potentials using a fully microscopic theory, we now formulate the two-body scattering problem for hybrid excitons within the $T$-matrix approach. Our approach has multiple distinct advantages over the use of the Born approximation. In particular, as we discuss, it is guaranteed to satisfy conservation of probability in a scattering process, and it reproduces the known universal low-energy scattering behavior~\cite{Landau2013quantum}. 

\subsection{Scattering integral equation for hybrid excitons}
\label{subsec:Scattering integral equation for hybrid excitons}
Since the system is Galilean invariant, we can work in the center-of-mass frame. We thus restrict our attention to two-exciton states with zero total momentum, so that the incoming (outgoing) excitons carry momenta $\pm\Q$ ($\pm\Q'$). 

In the basis consisting of DX and IX operators, the relevant two-particle states are then conveniently written in index notation. For a fixed valley sector $\eta$, the symmetric states are
\begin{align}
    &\ket{1;\Q,\eta} = \hat{x}^{\dagger}_{\Q,\eta}\hat{x}^{\dagger}_{-\Q,\eta}\ket{0}, \quad
    \ket{2;\Q,\eta} = \hat{y}^{\dagger}_{\Q,\eta}\hat{y}^{\dagger}_{-\Q,\eta}\ket{0}, \notag \\[0.2cm] 
    &\ket{3;\Q,\eta} = \frac{1}{\sqrt{2}} \left( \hat{x}^{\dagger}_{\Q,\eta}\hat{y}^{\dagger}_{-\Q,\eta} + \hat{y}^{\dagger}_{\Q,\eta}\hat{x}^{\dagger}_{-\Q,\eta} \right) \ket{0},
    \label{eq:two-particle states sym}
\end{align}
together with the antisymmetric DX-IX combination
\begin{equation}
    \ket{4;\Q,\eta} = \frac{1}{\sqrt{2}} \left( \hat{x}^{\dagger}_{\Q,\eta}\hat{y}^{\dagger}_{-\Q,\eta} - \hat{y}^{\dagger}_{\Q,\eta}\hat{x}^{\dagger}_{-\Q,\eta} \right) \ket{0}.
    \label{eq:two-particle states asym}
\end{equation}
For scattering from rotationally symmetric potentials at zero center-of-mass momentum, the $T$-matrix equation decouples into partial-wave channels labeled by the angular momentum quantum number $l$. Moreover, even-$l$ partial waves couple only to the symmetric sector in Eq.~\eqref{eq:two-particle states sym}, while odd-$l$ couple to the antisymmetric sector in Eq.~\eqref{eq:two-particle states asym}. In the low-energy regime, scattering is dominated by the $s$-wave ($l=0$) channel, and we therefore restrict to the symmetric sector. For brevity, we suppress the angular momentum quantum number in the following.

The strength of the interaction is quantified by the scattering $T$ matrix, which generalizes the Born approximation to arbitrary numbers of scattering events. In other words, we seek to calculate the two-exciton matrix elements of the operator
\begin{equation}
    \hat T=\hat V+\hat V\frac1{E-\hat H_0+i0}\hat T,
\end{equation}
where the Born approximation consists of keeping only the first term $\hat V$ on the right-hand side. Here, the infinitesimal positive imaginary part $+i0$ shifts the energy poles slightly into the lower half plane, as is appropriate for scattering events~\cite{sakurai2020}.

Projecting the operators $\hat{V}$ and $\hat{T}$ onto the two-particle basis $\ket{i;\Q,\eta}$, we define the $s$-wave potential 
\begin{equation}
    V_{ii}(Q',Q) \delta_{ij} = \frac{1}{2}\int^{2\pi}_{0}\!\frac{d\theta_{\Q'\Q}}{2\pi} \bra{i;\Q',\eta}\hat{V}\ket{j;\Q,\eta} \, , 
    \label{eq:partial wave potential matrix element}
\end{equation}
and similarly the $s$-wave $T$-matrix element
\begin{equation}
    T^{(\eta)}_{ij}(Q',Q;E) = \frac{1}{2}\int^{2\pi}_{0}\!\frac{d\theta_{\Q'\Q}}{2\pi}\bra{i;\Q',\eta}\hat{T}\ket{j;\Q,\eta} ,
    \label{eq:partial wave T-matrix element}
\end{equation}
where $\theta_{\Q'\Q}$ denotes the relative angle between $\Q'$ and $\Q$, and the factor $1/2$ removes the symmetry factor arising from scattering between identical particles. The $s$-wave $T$-matrix elements between the outgoing and incoming two-particle states then satisfy the coupled integral equation~\cite{Nakano2026}
\begin{align}
    &T^{(\eta)}_{ij}(Q',Q;E) = V_{ii}(Q',Q)\delta_{ij} \notag \\ 
    &+ \sum^{3}_{n=1}\int^{\infty}_{0}\!\frac{q\,dq}{2\pi} V_{ii}(Q',q)G^{(\eta)}_{in}(q,E)T^{(\eta)}_{nj}(q,Q;E).
    \label{eq:partial wave scattering integral equation}
\end{align}
A diagrammatic representation of the integral equation is shown in Fig.~\ref{fig:T-matrix equation}. In our description, the pseudopotentials are chosen such that their matrix elements reproduce the microscopic Born approximation for exciton-exciton scattering in the zero-momentum limit. The first term on the right-hand side of Eq.~\eqref{eq:partial wave scattering integral equation} therefore contains the Born contribution, whereas the second term resums repeated scattering processes and channel mixing to all orders.

Using the correspondence between the basis states [Eq.~\eqref{eq:two-particle states sym}] and the pseudopotentials in the potential operator [Eq.~\eqref{eq:potential operator}], we identify $V_{11}=V_\text{DX-DX}$, $V_{22}=V_\text{IX-IX}$, and $V_{33}=V_\text{DX-IX}$. Within our model, the interaction does not modify the internal structure of the excitons, and thus the potential is diagonal in the channel index. 

The two-particle propagator $G$ introduced in Eq.~\eqref{eq:partial wave scattering integral equation} is defined via
\begin{equation}
    G^{(\eta)}_{in}(q;E)
    = \bra{i;\q,\eta}\frac{1}{(E+i0)\mathbb{1}-\hat{H}^{(\eta)}_{0}}\ket{n;\q,\eta}.
    \label{eq:Green's function}
\end{equation}
Here $\hat{H}^{(\eta)}_0$ is the non-interacting hybrid-exciton Hamiltonian in Eq.~\eqref{eq:non-interacting Hamiltonian hX}, and $\mathbb{1}$ is the identity operator, with completeness relation $\mathbb{1} = \frac{1}{2}\sum_{\q}\sum^{4}_{i=1} \ket{i;\q,\eta}\bra{i;\q,\eta}$. Although the interaction potential is diagonal as in Eq.~\eqref{eq:partial wave scattering integral equation}, the matrix $G^{(\eta)}_{in}$ is generally not diagonal because $\hat H_0$ hybridizes DX and IX at the single-particle level, leading to channel mixing during free propagation between successive scattering events. The explicit form of the propagator is given in Appendix~\ref{appx:two-particle propagator}.

Finally, to obtain the hybrid-exciton $T$ matrix, we introduce the two-hybrid-exciton state at zero total momentum
\begin{equation}
    \ket{A,B;\Q,\eta} = \hat{A}^{\dagger}_{\Q,\eta}\hat{B}^{\dagger}_{-\Q,\eta}\ket{0},
    \label{eq:two-particle states hX}
\end{equation}
where $A,B\in\{L,U\}$ label the lower and upper hybrid-exciton modes, and the corresponding operators are defined in Eq.~\eqref{eq:hybrid exciton operators}. Projecting the $s$-wave exciton $T$-matrix in the symmetric basis [Eq.~\eqref{eq:partial wave scattering integral equation}] onto the two-hybrid-exciton state yields
\begin{align}
    &T^{(\eta)}_{AB}(Q',Q;E) = \frac{2}{1+\delta_{AB}}\sum^{3}_{i,j=1} \bra{A,B;\Q',\eta}\ket{i;\Q',\eta} \notag \\ 
    &\quad \times \bra{j;\Q,\eta} \ket{A,B;\Q,\eta} T^{(\eta)}_{ij}(Q',Q;E),
    \label{eq:hybrid exciton T matrix}
\end{align}
where the prefactor accounts for the indistinguishability of the two hybrid excitons with $\delta_{AB}$ denoting the Kronecker delta. The overlap coefficients $\braket{i;\Q,\eta}{A,B;\Q,\eta}$ follow directly from the linear transformation in Eq.~\eqref{eq:hybrid exciton operators} and are given in Appendix~\ref{appx:two-particle propagator}. 

The collision energy $E$ denotes the total two-body energy in the center-of-mass frame.
For on-shell elastic scattering with an incoming state $(A,B)$, we set
\begin{equation}
    E = E^{A}_{\Q,\eta} + E^{B}_{\Q,\eta},
    \label{eq:on-shell energy}
\end{equation}
where $E^{A}_{\Q,\eta}$ is the hybrid-exciton dispersion in valley sector $\eta$ as defined in Eq.~\eqref{eq:hybrid exciton dispersions} with the $+$ sign for $A=U$ and the $-$ sign for $A=L$. When evaluating observables from Eq.~\eqref{eq:hybrid exciton T matrix}, energy conservation further implies $|\Q'|=|\Q|$ for on-shell scattering. Off-shell values $|\Q'|\neq |\Q|$ enter only virtually, internally in the integral equation~\eqref{eq:partial wave scattering integral equation}.

For further convenience, we define the dimensionless scattering amplitude for hybrid excitons as
\begin{equation}
    f^{(\eta)}_{AB}(Q) = m_{\text{X}}\,T^{(\eta)}_{AB}(Q,Q;E),
    \label{eq:hybrid exciton scattering amplitude}
\end{equation}
evaluated at the on-shell collision energy.

\subsection{Exciton $T$ matrix}
We first consider the exciton $T$ matrices in the absence of hole tunneling. Setting $t=0$ in the $s$-wave scattering integral equation in Eq.~\eqref{eq:partial wave scattering integral equation} removes the single-particle DX-IX hybridization, so that the two-exciton scattering problem reduces to collisions between excitons of fixed internal structure. Within the index notation introduced in Sec.~\ref{subsec:Scattering integral equation for hybrid excitons}, the DX-DX, IX-IX and DX-IX $T$-matrices correspond to the diagonal matrix elements $T_{11} = T_\text{DX-DX}$, $T_{22} = T_\text{IX-IX}$ and $T_{33} = T_\text{DX-IX}$, respectively.

To gain insight into the evolution with momentum, it is useful to recall the universal form of the elastic $s$-wave scattering amplitude in 2D~\cite{Landau2013quantum,Levinsen2015review}
\begin{equation}
    f_\text{X-X}(Q) = \frac{-4}{\cot\delta_{0}(Q)-i}.
\end{equation}
In particular, in the absence of inelastic decay channels in the scattering, i.e., for real phase shifts $\delta_{0}(Q)$, conservation of probability implies the elastic unitarity condition
\begin{equation}
    (\text{Re}\,f_\text{X-X})^2+(\text{Im}\,f_\text{X-X}+2)^2=4.
    \label{eq:elastic unitarity}
\end{equation}
This leads to the bounds $-2 \le \text{Re}\,f_\text{X-X} \le 2$ and $-4 \le \text{Im}\,f_\text{X-X} \le 0$. With our conventions, the dimensionless $s$-wave amplitude is proportional to the on-shell $T$ matrix via 
\begin{equation}
    f_\text{X-X}(Q) = m_{\text{X}}\,T_\text{X-X}(Q,Q;E),
\end{equation}
so the same unitarity constraints control the relative magnitudes of the real and imaginary parts of the $T$ matrix.

\begin{figure}[tp]
    \begin{minipage}{0.48\textwidth}
    \centering
    \includegraphics[width=\textwidth]{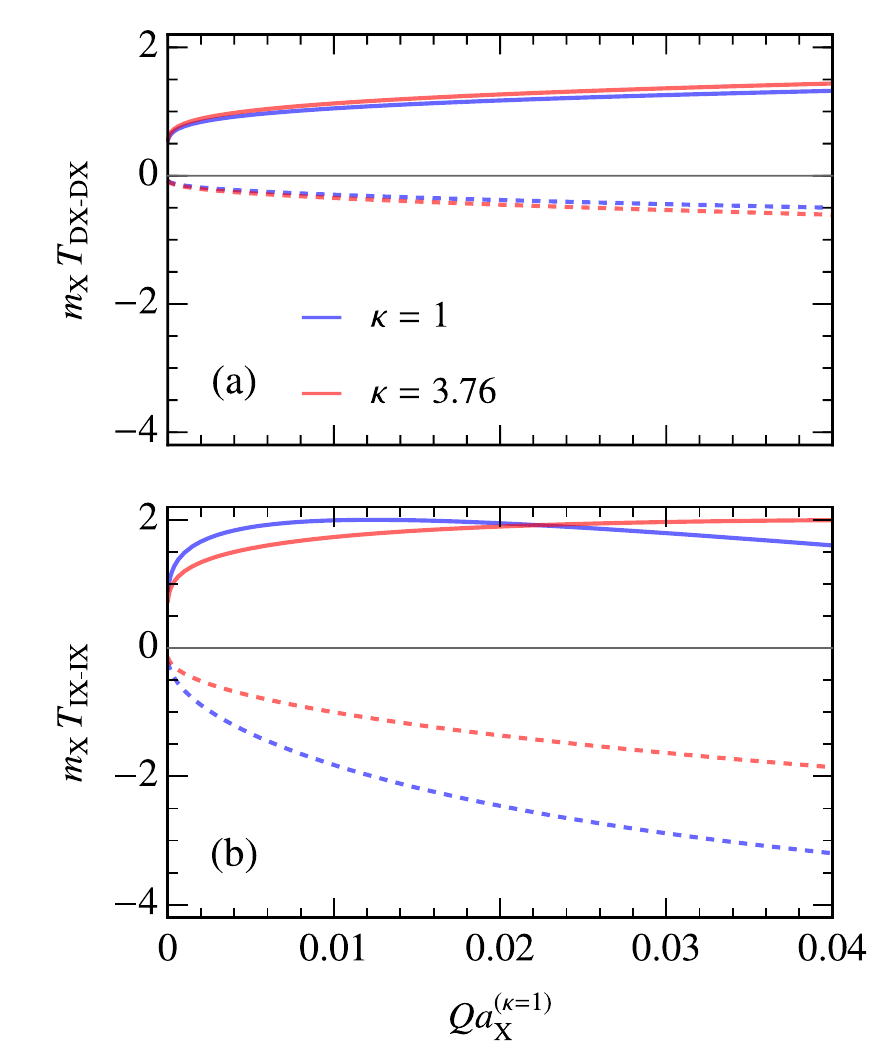}
    \end{minipage}
    \caption{Real (solid) and imaginary (dashed) parts of the exciton $T$ matrices as a function of momentum for (a) DX-DX and (b) IX-IX scattering. Blue and red lines show results for vacuum ($\kappa=1$) and hBN ($\kappa=3.76$) encapsulation, respectively. We assume equal electron and hole masses, and we use a fixed layer separation and polarizability (Table~\ref{tab:parameters}). The panels are plotted in terms of the fixed reference scale $a_\text{X}^{(\kappa=1)}$. We do not show the DX-IX scattering, since we find that it is nearly identical to DX-DX scattering.}
    \label{fig:TXX}
\end{figure}

\begin{figure}[tp]
    \begin{minipage}{0.48\textwidth}
    \centering
    \includegraphics[width=\textwidth]{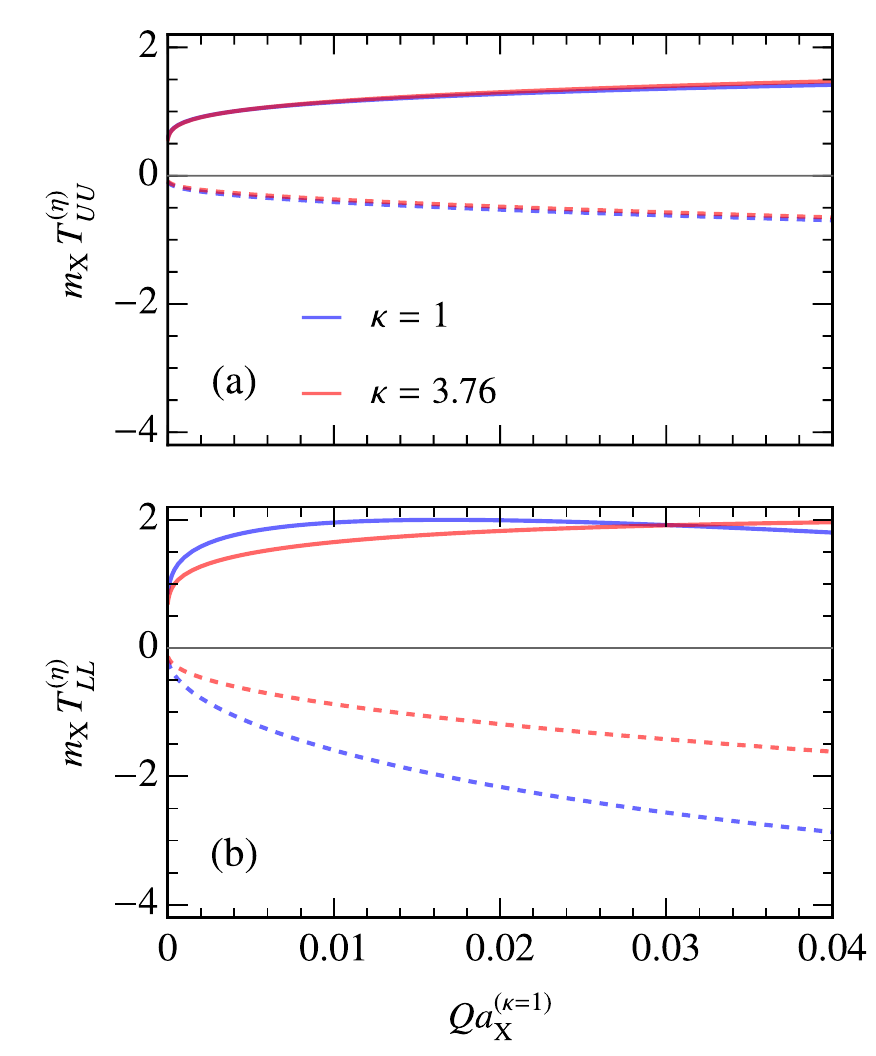}
    \end{minipage}
    \caption{Real (solid) and imaginary (dashed) parts of the hybrid exciton $T$ matrices as a function of momentum at $\Delta=0$ for (a) hX$^+$-hX$^+$ and (b) hX$^-$-hX$^-$ scattering. Blue and red lines correspond to vacuum ($\kappa=1$) and hBN ($\kappa=3.76$) encapsulation, respectively. We assume equal electron and hole masses, and we use a fixed layer separation and polarizability (Table~\ref{tab:parameters}) with the tunneling rate $t/\varepsilon^{(\kappa=1)}_\text{X} = 0.17$, and IX detuning $\delta_\text{IX}/\varepsilon^{(\kappa=1)}_\text{X} = -0.17$. The panels are plotted in terms of the fixed reference scale $a_\text{X}^{(\kappa=1)}$.}
    \label{fig:ThXhX}
\end{figure}

Figure~\ref{fig:TXX} shows the real (solid) and imaginary (dashed) parts of the on-shell $s$-wave $T$ matrices as functions of the relative momentum. A clear qualitative difference emerges between DX-DX and IX-IX scattering. For IX-IX, both the real and imaginary parts of the $T$ matrix evolve rapidly with $\Q$ and approach their unitarity-limited values already at comparatively small momenta. This reflects a rapid increase of the $s$-wave phase shift $\delta_{0}(Q)$ at small $\Q$, which we attribute to the long-range repulsive dipole-dipole interaction between IXs. By contrast, DX-DX interactions are effectively shorter-ranged, and the associated phase shift evolves more gradually with momentum. The same applies to the DX-IX interactions (not shown), which are also short ranged and hence quantitatively similar to those of the DX-DX case. Thus, in the regime of low-momentum collisions, $Q\ll a_\text{X}^{-1}$, typical for experiments at cryogenic temperatures (see Appendix~\ref{appx:estimate of the thermally relevant momentum}), we find that the IX-IX interactions dominate.

Furthermore, the dielectric environment affects the two channels in distinct ways. For DX-DX scattering, we find that the magnitude of the $T$ matrix is typically larger for hBN than for vacuum. This behavior can be attributed to the larger dielectric screening, which reduces the DX binding energy and increases the exciton size, thereby enhancing the effective interaction range and strengthening low-energy scattering. For IX-IX scattering, this trend is reversed, where the vacuum case exhibits faster phase evolution and a larger interaction strength than the hBN case. This is consistent with the fact that the long-range dipole-dipole repulsion is directly suppressed by dielectric screening, so increasing $\kappa$ weakens the dipolar tail and reduces the low-momentum scattering strength.

It is useful to compare our results with the Born approximation that is conventionally applied to the scattering of excitons. In that case, the interaction constant is typically evaluated at zero relative momentum~\cite{Ciuti1998,Tassone1999}. However, we note that, formally, the exciton-exciton interaction must vanish at zero momentum, as it does for any short-range interaction in a 2D geometry~\cite{Landau2013quantum}. Indeed, this is the case for our non-perturbative results in Fig.~\ref{fig:TXX}, where we have $T\sim 1/\log(1/Q)$ for small $Q$ which vanishes logarithmically as $Q\to0$. Of course, experiments measure exciton interaction effects, and the resolution to this apparent discrepancy is that the collision energy never exactly vanishes. Instead, it can be set by temperature, the exciton density, the linewidth, or the finite size of the system. 

\begin{figure}[tp]
    \begin{minipage}{0.48\textwidth}
    \centering
    \includegraphics[width=\textwidth]{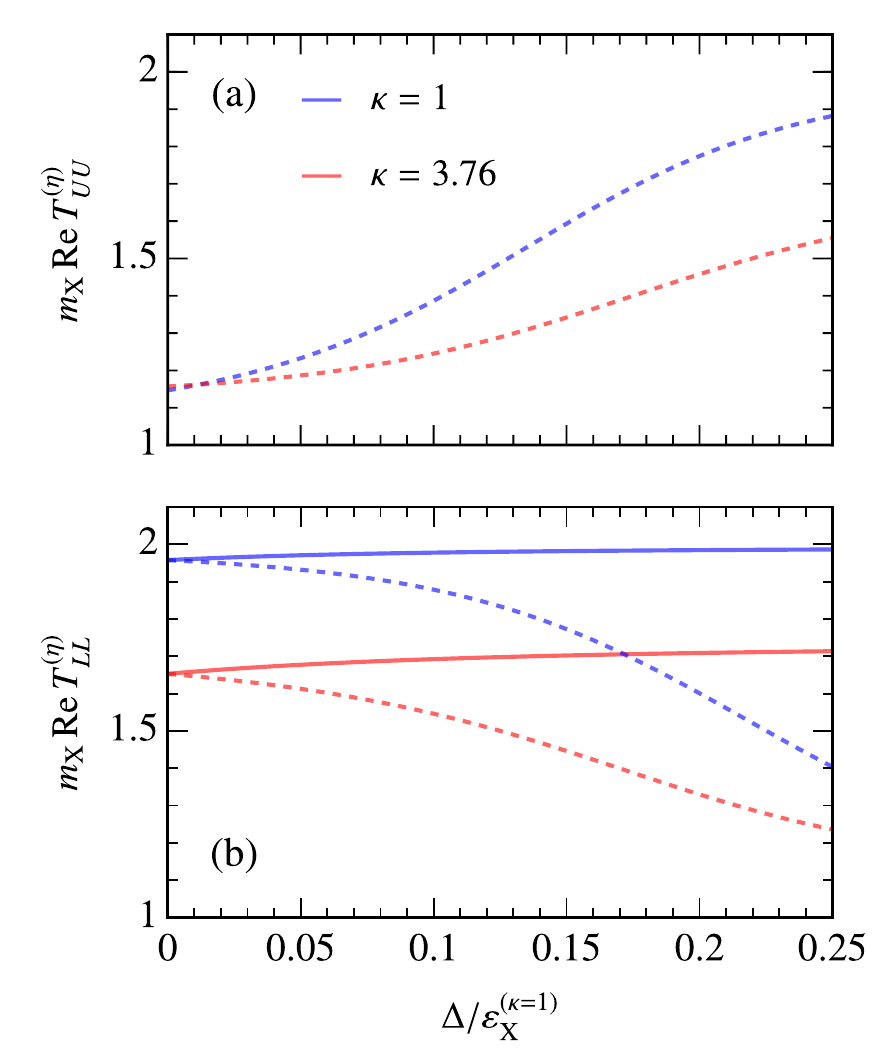}
    \end{minipage}
    \caption{Real part of the hybrid-exciton $T$ matrices as a function of Stark shift evaluated at fixed momentum $Qa^{(\kappa=1)}_\text{X}=0.01$ for (a) hX$^+$-hX$^+$ and (b) hX$^-$-hX$^-$ scattering. Blue and red lines correspond to $\kappa=1$ (vacuum) and $\kappa=3.76$ (hBN), respectively. For each $\kappa$, solid and dashed lines show the $\eta=1$ and $\eta=2$ sectors. In all panels, we take equal electron and hole masses, a fixed layer separation and polarizability (Table~\ref{tab:parameters}), tunneling rate $t/\varepsilon^{(\kappa=1)}_\text{X} = 0.17$, and IX detuning $\delta_\text{IX}/\varepsilon^{(\kappa=1)}_\text{X} = -0.17$.}
    \label{fig:ThXhX Stark}
\end{figure}

\begin{figure}[tp]
    \begin{minipage}{0.48\textwidth}
    \centering
    \includegraphics[width=\textwidth]{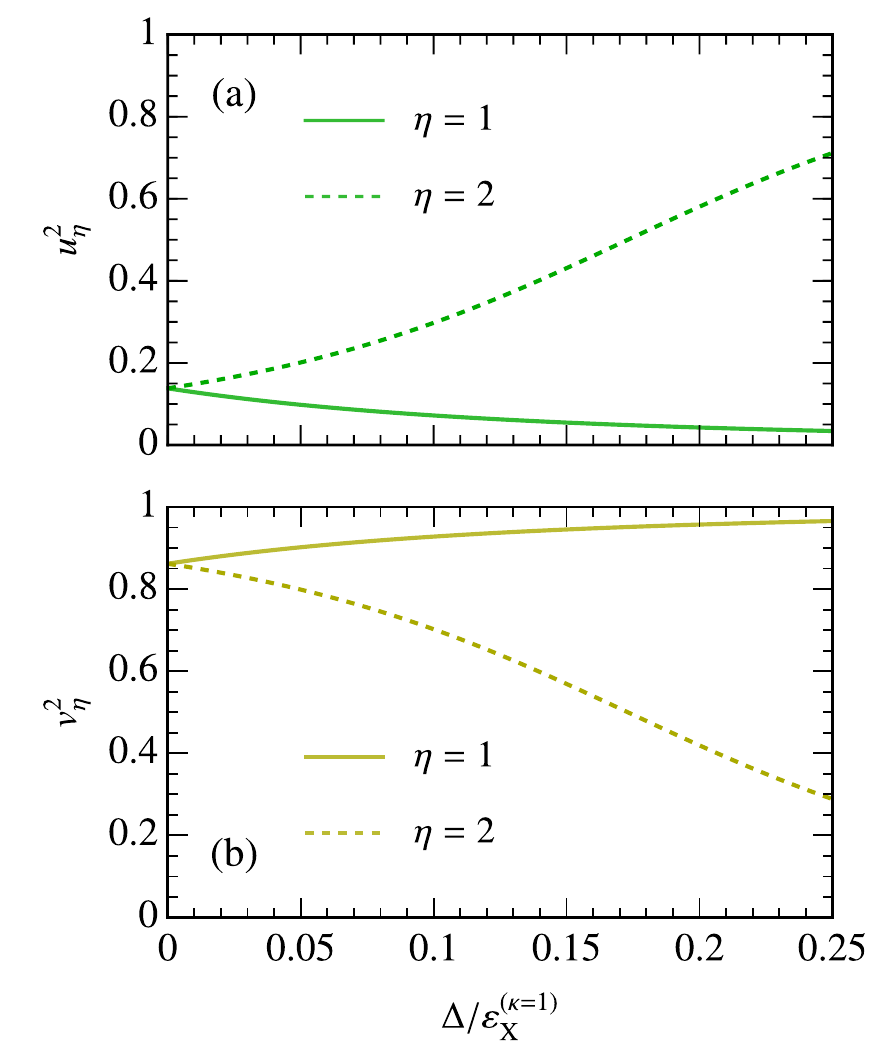}
    \end{minipage}
    \caption{Transformation coefficients (a) $u_{\eta}^2$ and (b) $v_{\eta}^2$ as a function of Stark shift. For the hX$^{+}$ mode, the IX fraction is $u^{2}_\eta$, whereas for the hX$^{-}$ mode, the IX fraction is $v^{2}_\eta$. Solid and dashed lines correspond to the $\eta=1$ and $\eta=2$ sectors, respectively. For equal DX and IX masses, the coefficients are momentum independent. We use a tunneling rate $t/\varepsilon^{(\kappa=1)}_\text{X} = 0.17$ and IX detuning $\delta_\text{IX}/\varepsilon^{(\kappa=1)}_\text{X} = -0.17$.}
    \label{fig:transformation coeff}
\end{figure}

\subsection{Hybrid exciton $T$ matrix}
We now turn to hybrid exciton scattering by explicitly including the DX-IX hybridization in the scattering integral equation. In the following, we focus on elastic scattering within a given mode and valley, namely lower-lower (hX$^-$-hX$^-$) and upper-upper (hX$^+$-hX$^+$) hybrid exciton scattering for fixed $\eta$. As discussed previously, the corresponding hybrid-exciton $T$ matrices are defined by projecting the coupled-channel $T$ matrix [Eq.~\eqref{eq:partial wave scattering integral equation}] in the bare two-exciton basis onto the hybrid-exciton two-particle states [Eq.~\eqref{eq:two-particle states hX}]. More specifically, we obtain the hybrid exciton $T$ matrix from Eq.~\eqref{eq:hybrid exciton T matrix} by choosing $A=B=L$ for hX$^-$-hX$^-$ scattering or $A=B=U$ for hX$^+$-hX$^+$ scattering. All virtual processes mediated by the underlying DX and IX channels are retained in the exciton $T$ matrix $T_{ij}$, while the hybrid exciton $T$ matrix $T_{AB}$ captures the effective scattering amplitude between the physical hybrid eigenmodes.

In Fig.~\ref{fig:ThXhX}, we show the on-shell hybrid-exciton $T$ matrices evaluated at zero applied out-of-plane electric field. In this case, the two sectors $\eta=1,2$ are degenerate and therefore yield identical scattering amplitudes. For a 2H MoS$_2$ homobilayer, the bare IX lies at lower energy than the bare DX. As a result, the hX$^+$ is predominantly DX-like while the hX$^-$ is predominantly IX-like at $\Delta=0$. Correspondingly, in Fig.~\ref{fig:ThXhX}(a), the hX$^+$-hX$^+$ $T$ matrix closely resembles the bare DX-DX result in Fig.~\ref{fig:TXX}(a), whereas the hX$^-$-hX$^-$ $T$ matrix in Fig.~\ref{fig:ThXhX}(b) follows the qualitative behavior of the bare IX-IX $T$ matrix in Fig.~\ref{fig:TXX}(b), including the rapid approach to the unitarity bound at comparatively small momenta due to the long-range dipolar repulsion.

Importantly, turning on the DX-IX hybridization does not fundamentally violate elastic unitarity in Eq.~\eqref{eq:elastic unitarity}. Instead, the on-shell hybrid-exciton scattering amplitude lies within the same unitarity circle
\begin{equation}
    (\text{Re}\,f^{(\eta)}_{AB})^2+(\text{Im}\,f^{(\eta)}_{AB}+2)^2 \le 4,
\end{equation}
where the $\le$ sign is due to the presence of additional scattering channels.
This behavior is qualitatively different from microcavity polariton-polariton scattering, which can likewise be formulated in a coupled-channel approach~\cite{Bleu2020,Nakano2026}. There, hybridization of an exciton with a cavity photon yields an extremely small polariton effective mass and places the two-polariton energy at small momenta below the bare two-exciton continuum. When polariton-polariton scattering is mapped onto an exciton-only description, this corresponds to probing the excitonic interaction at an effectively negative collision energy~\cite{Bleu2020,Nakano2026}, for which the simple single-channel elastic unitarity relation is not directly applicable.

\subsection{Stark shift}
A key control parameter for hybrid-exciton scattering is the electrically induced Stark shift, which tunes the relative DX-IX detuning and thus the hybridization between the two modes. Figure~\ref{fig:ThXhX Stark} highlights this by showing the real part of the hybrid-exciton $T$-matrices, evaluated at fixed momentum, as a function of Stark shift for both $\kappa=1$ and $\kappa=3.76$. The observed behaviors can be understood from the DX and IX weights of the hybrid-exciton modes, as shown in Fig.~\ref{fig:transformation coeff}. Specifically, for the hX$^{+}$ mode, the IX fraction is $u^{2}_\eta$, whereas for the hX$^{-}$ mode, the IX fraction is $v^{2}_\eta$.

We first consider hX$^{+}$-hX$^{+}$ scattering in Fig.~\ref{fig:ThXhX Stark}(a). As the Stark shift increases, the IX in the $\eta=2$ sector shifts upward and the hX$^{+}$ mode becomes progressively more indirect excitonic, as shown by the increasing IX fraction $u^{2}_\eta$ in Fig.~\ref{fig:transformation coeff}(a). Correspondingly, the interaction strength increases with $\Delta$ owing to the enhanced dipolar character of the hX$^{+}$ mode. We do not show the $\eta=1$ sector in Fig.~\ref{fig:ThXhX Stark}(a), since for the positive Stark shifts it lies outside the regime where the two-mode coupled-oscillator model accurately describes the microscopic spectrum, as shown in Fig.~\ref{fig:coupled oscillator}(a).

Turning to hX$^{-}$-hX$^{-}$ scattering in Fig.~\ref{fig:ThXhX Stark}(b), the $\eta=1$ sector acquires a larger IX fraction $v^{2}_{\eta}$ with increasing Stark shift, as shown in Fig.~\ref{fig:transformation coeff}(b), and therefore exhibits an increasing interaction strength. By contrast, the $\eta=2$ sector becomes more direct excitonic and shows a reduced interaction strength. This behavior qualitatively agrees with Ref.~\cite{Federolf2025}, which also reported a stronger field-induced energy blueshift for the $\eta=1$ branch of the hX$^{-}$ mode than the $\eta=2$ branch. More quantitative agreement may require including intervalley scattering processes, which are expected to provide an additional attractive contribution due to antiparallel dipole-dipole interactions between IXs in opposite valleys, as well as density-dependent energy shifts and additional branches associated with excited exciton states predicted by the microscopic coupled \sch equation in Eq.~\eqref{eq:coupled schrodinger equation}.

Overall, Fig.~\ref{fig:ThXhX Stark} demonstrates that electrical control of the DX-IX detuning provides a direct means to continuously vary the hybrid-exciton scattering strength between the DX-dominated and IX-dominated limits.

\section{Conclusions}
\label{sec:conclusions}
In this work, we developed a microscopic description of the direct and indirect excitons in a MoS$_2$ homobilayer and used it to evaluate the effective exciton-exciton interaction strengths within the Born approximation. Building on these microscopic results, we established effective pseudopotentials that faithfully capture the low-energy two-exciton physics, which in turn enabled a fully non-perturbative two-body scattering calculation for hybrid excitons using a $T$-matrix approach. We showed that the long-range dipolar interactions between IXs exhibit a more rapid evolution of the interaction strength at comparatively small momenta relative to short-range interactions between DXs. 

Upon including the DX-IX hybridization, we further demonstrated that hybrid-exciton scattering interpolates between direct excitonic and indirect excitonic characters and can be electrically tuned via the Stark shift. Electrical control of exciton-exciton interactions provides a versatile knob for accessing distinct interaction regimes without changing the sample geometry. In particular, it allows one to enhance or suppress effective repulsion, and thereby tailor conditions for nonlinear transport and interaction-driven dynamics in excitonic and polaritonic devices~\cite{Sanvitto2016,Liew2023}.

Our coupled-channel approach can be naturally extended to more complex hybridization scenarios such as dipolariton scattering in microcavities~\cite{Togan2018,Datta2022,Louca2023,Xiang2026}. It also applies to other dressed-state platforms, such as Rabi-coupled cold-atom systems where interactions between internal-state superpositions can be tuned and treated within the multichannel $T$-matrix description~\cite{Bleu2025,Zulli2025}.

In this work, we have focused on excitons with the same underlying electron spins, which can be excited using circularly polarized photons. The extension to opposite spins is of great interest since it can feature strong enhancement of interactions due to Feshbach resonances associated with biexciton bound states. Accessing these requires a tunable collision energy, which can be realized in the context of exciton-polariton scattering due to the ability to tune the photon detuning~\cite{Wouters2007,Takemura2014,Bleu2020,Tan2023}. Exploring the use of the Stark effect in hybrid excitons to achieve such resonantly enhanced interactions offers a compelling avenue for future research.

\acknowledgments 
We thank Olivier Bleu and Brendan Mulkerin for valuable discussions.
We acknowledge support from the Australian Research Council Centre of Excellence in Future Low-Energy Electronics Technologies (CE170100039). JL and MMP are also supported through Australian Research Council Discovery Project DP240100569 and Future Fellowship FT200100619, respectively.

\onecolumngrid
\appendix

\section{Two-particle propagator}
\label{appx:two-particle propagator}

For a fixed sector $\eta$, the non-interacting two-particle propagator in Eq.~\eqref{eq:Green's function} is obtained by a spectral decomposition over the two-hybrid-exciton state in Eq.~\eqref{eq:two-particle states hX}. It takes the form
\begin{equation}
    G^{(\eta)}_{in}(\q;E) = \sum_{A,B} \frac{\bra{i;\q} \ket{A,B;\q}\bra{A,B;\q}\ket{n;\q}}{E-E^{A}_{\q,\eta}-E^{B}_{\q,\eta}+i0},
    \label{eq:two-particle propagator appx}
\end{equation}
where $E^{\pm}_{\q,\eta}$ are the single-particle hybrid-exciton dispersions in Eq.~\eqref{eq:hybrid exciton dispersions} with the $+$ sign for $A=U$ and the $-$ sign for $A=L$. Equation~\eqref{eq:two-particle propagator appx} shows that off-diagonal elements in $i,n$ arise from the overlap between bare and hybrid two-particle states. The corresponding overlap coefficients follow directly from the symmetrized two-particle basis states in Eq.~\eqref{eq:two-particle states sym} together with the linear transformation in Eq.~\eqref{eq:hybrid exciton operators}. Writing the hX$^{-}$ and hX$^{+}$ modes in terms of their DX and IX components, the overlap between a bare two-exciton state and a hybrid two-particle state follows
\begin{equation}
    \bra{i;\q}\ket{A,B;\q} =
    \begin{dcases}
    X^{A}_{\q,\eta}X^{B}_{\q,\eta}, \quad &\text{for} \quad i = 1, \\[0.2cm] 
    Y^{A}_{\q,\eta}Y^{B}_{\q,\eta}, \quad &\text{for} \quad i = 2, \\[0.2cm] 
    \frac{1}{\sqrt{2}} \left( X^{A}_{\q,\eta}Y^{B}_{\q,\eta}+Y^{A}_{\q,\eta}X^{B}_{\q,\eta} \right), \quad &\text{for} \quad i = 3,
    \end{dcases}
    \label{eq:overlap coefficients appx}
\end{equation}
with $X^{-}_{\q,\eta} = Y^{+}_{\q,\eta} = u_{\q,\eta}$ and $Y^{-}_{\q,\eta} = -X^{+}_{\q,\eta} = v_{\q,\eta}$.

Substituting Eq.~\eqref{eq:overlap coefficients appx} into Eq.~\eqref{eq:two-particle propagator appx} and summing over the relevant two-particle hybrid configurations yields the explicit matrix elements
\begin{align}
    G^{(\eta)}_{11}(\q;E) &= \frac{u^{4}_{\q,\eta}}{E-2E^{-}_{\q,\eta}+i0} + \frac{2u^{2}_{\q,\eta}v^{2}_{\q,\eta}}{E-E^{-}_{\q,\eta}-E^{+}_{\q,\eta}+i0} + \frac{v^{4}_{\q,\eta}}{E-2E^{+}_{\q,\eta}+i0}, \notag \\[0.2cm] 
    G^{(\eta)}_{12}(\q;E) = G^{(\eta)}_{21}(\q;E) &= \frac{u^{2}_{\q,\eta}v^{2}_{\q,\eta}}{E-2E^{-}_{\q,\eta}+i0} - \frac{2u^{2}_{\q,\eta}v^{2}_{\q,\eta}}{E-E^{-}_{\q,\eta}-E^{+}_{\q,\eta}+i0} + \frac{u^{2}_{\q,\eta}v^{2}_{\q,\eta}}{E-2E^{+}_{\q,\eta}+i0}, \notag \\[0.2cm] 
    G^{(\eta)}_{13}(\q;E) = G^{(\eta)}_{31}(\q;E) &= -\frac{\sqrt{2}u^{3}_{\q,\eta}v_{\q,\eta}}{E-2E^{-}_{\q,\eta}+i0} + \frac{\sqrt{2}u_{\q,\eta}v_{\q,\eta}(u^{2}_{\q,\eta}-v^{2}_{\q,\eta})}{E-E^{-}_{\q,\eta}-E^{+}_{\q,\eta}+i0} + \frac{\sqrt{2}u_{\q,\eta}v^{3}_{\q,\eta}}{E-2E^{+}_{\q,\eta}+i0}, \notag \\[0.2cm] 
    G^{(\eta)}_{22}(\q;E) &= \frac{v^{4}_{\q,\eta}}{E-2E^{-}_{\q,\eta}+i0} + \frac{2u^{2}_{\q,\eta}v^{2}_{\q,\eta}}{E-E^{-}_{\q,\eta}-E^{+}_{\q,\eta}+i0} + \frac{u^{4}_{\q,\eta}}{E-2E^{+}_{\q,\eta}+i0}, \notag \\[0.2cm] 
    G^{(\eta)}_{23}(\q;E) = G^{(\eta)}_{32}(\q;E) &= -\frac{\sqrt{2}u_{\q,\eta}v^{3}_{\q,\eta}}{E-2E^{-}_{\q,\eta}+i0} - \frac{\sqrt{2}u_{\q,\eta}v_{\q,\eta}(u^{2}_{\q,\eta}-v^{2}_{\q,\eta})}{E-E^{-}_{\q,\eta}-E^{+}_{\q,\eta}+i0} + \frac{\sqrt{2}u^{3}_{\q,\eta}v_{\q,\eta}}{E-2E^{+}_{\q,\eta}+i0}, \notag \\[0.2cm] 
    G^{(\eta)}_{33}(\q;E) &= \frac{2u^{2}_{\q,\eta}v^{2}_{\q,\eta}}{E-2E^{-}_{\q,\eta}+i0} + \frac{(u^{2}_{\q,\eta}-v^{2}_{\q,\eta})^{2}}{E-E^{-}_{\q,\eta}-E^{+}_{\q,\eta}+i0} + \frac{2u^{2}_{\q,\eta}v^{2}_{\q,\eta}}{E-2E^{+}_{\q,\eta}+i0}.
    \label{eq:hybrid exciton propagator appx}
\end{align}
The denominators correspond to propagation through two hX$^{-}$ modes, one hX$^{-}$ and one hX$^{+}$ modes, or two hX$^{+}$ modes, respectively. The numerators encode the weights with which each hybrid configuration projects onto a given bare two-exciton channel. In particular, the off-diagonal matrix elements arise from the hybridization between DX and IX components and vanish in the absence of tunneling, in which case one recovers a diagonal propagator in the bare channel basis.

\section{Estimate of the thermally relevant momentum}
\label{appx:estimate of the thermally relevant momentum}
Here, we provide an estimate of the thermal momentum for two-exciton scattering relevant to cryogenic experiments. For completeness, we explicitly write the Planck constant $\hbar$ throughout this section. For elastic scattering between two excitons at zero center-of-mass momentum, the collision energy is
\begin{equation}
    E_\text{coll}(Q)
    = \frac{\hbar^{2}Q^{2}}{m_\text{X}},
\end{equation}
where $m_\text{X}$ denotes the exciton mass. Assuming that the excitons are thermally distributed according to the Boltzmann distribution, the characteristic collision momentum at temperature $T$ can be estimated by relating the collision energy to the thermal energy scale
\begin{equation}
    E_{\text{coll}}(Q_\text{th}) \sim k_\text{B}T,
\end{equation}
with $k_\text{B}$ the Boltzmann constant. This yields the thermal momentum
\begin{equation}
    \hbar Q_\text{th} \sim \sqrt{m_\text{X}k_\text{B}T},
    \label{eq:thermal momentum appx}
\end{equation}
which can be interpreted as a momentum scale around which thermally populated two-exciton scattering processes are expected to occur.

In the present work, we assume the equal electron and hole masses $m_\text{e} = m_\text{h} = 0.5m_{0}$, where $m_{0}$ denotes the bare electron mass. The corresponding exciton mass is therefore $m_\text{X} = m_\text{e} + m_\text{h} = m_{0}$. For a cryogenic temperature $T = 4~\text{K}$, Eq.~\eqref{eq:thermal momentum appx} then gives
\begin{equation}
    Q_\text{th} \approx 0.0673~\text{nm}^{-1}.
\end{equation}
Using the reference exciton size $a_\text{X}^{(1)}=0.576~\text{nm}$, we obtain
\begin{equation}
    Q_\text{th} a_\text{X}^{(1)} \approx 0.0387.
\end{equation}
Thus, the thermally relevant momenta at cryogenic temperatures satisfy $Qa_\text{X}^{(1)} \ll 1$.

\twocolumngrid

\bibliography{condensed-matter}

\end{document}